\documentclass[aip, floatfix, amsmath, amssymb, reprint,]{revtex4-1}
\pdfoutput=1
\usepackage{graphicx}
\usepackage{dcolumn}
\usepackage{bm}
\usepackage[utf8]{inputenc}
\usepackage[T1]{fontenc}
\usepackage{braket}
\usepackage{amsmath,amsfonts,amssymb, xparse}
\usepackage{etoolbox}
\usepackage{xcolor}

\makeatletter
\def\@email#1#2{
 \endgroup
 \patchcmd{\titleblock@produce}
  {\frontmatter@RRAPformat}
  {\frontmatter@RRAPformat{\produce@RRAP{*#1\href{mailto:#2}{#2}}}\frontmatter@RRAPformat}
  {}{}
}
\makeatother

\begin{document}

\preprint{AIP/123-QED}

\title[Critical assessment of ML repulsive potentials for the DFTB method: a case study for pure Si]{Critical assessment of machine-learned repulsive potentials for the Density Functional based Tight-Binding method: a case study for pure silicon}

\author{D. Bissuel}

\affiliation{ 
Univ Lyon, Université Claude Bernard Lyon 1, CNRS, Institut Lumière Matière, F-69622 Villeurbanne, France
} 

\author{T. Albaret}
\affiliation{ 
Univ Lyon, Université Claude Bernard Lyon 1, CNRS, Institut Lumière Matière, F-69622 Villeurbanne, France
}
\author{T.A. Niehaus}
\affiliation{ 
Univ Lyon, Université Claude Bernard Lyon 1, CNRS, Institut Lumière Matière, F-69622 Villeurbanne, France
} 
\email{
thomas.niehaus@univ-lyon1.fr
}
\date{\today}

\begin{abstract}
We investigate the feasability of improving the semi-empirical density functional based tight-binding method (DFTB) through a general and transferable many-body repulsive potential for pure silicon using a common machine-learning framework. Atomic environments using atom centered symmetry functions fed into flexible high-dimensional neural-networks allow to overcome the limited pair potentials used until now, with the ability to train simultaneously on a large variety of systems. We achieve an improvement on bulk systems, with good performance on energetic, vibrational and structural properties. Contrarily, there are difficulties for clusters due to surface effects. To deepen the discussion, we also put these results into perspective with two fully machine-learned numerical potentials for silicon from the literature. This allows us to identify both the transferability of such approaches together with the impact of narrowing the role of machine-learning models to reproduce only a part of the total energy. 
\end{abstract}

\maketitle

\section{\label{sec:introduction}Introduction}

Within the landscape of atomistic simulation, we observed during the last decade the emergence and increasing importance of many machine-learning (ML) methods (see for instance Ref.~\onlinecite{bib:marques_review} or Fig. 1 of Ref.~\onlinecite{bib:Behler_ML_simu_ato}). This appeal is widely justified by their well-known flexibility, so much that we may find them nowadays in any given field under some form or another. In particular the so-called numerical potentials\cite{bib:behler_perspective} attracted the interest of a large audience by aiming to reach chemical accuracy for large structures. One can also apply such methodologies to compute specific observables different from the total energy\cite{bib:bartok_water_correct_DFT}. 

Here, we will focus on the approximate density functional theory (DFT) method  Density Functional based Tight-Binding (DFTB)\cite{bib:dftb1,bib:dftb2,bib:dftb3}, widely used to efficiently compute electronic and structural properties at a much lower cost than common first-principles schemes. In particular, the repulsive contribution to the total DFTB energy has gathered some attention. This short-range term embeds a lot of complexity and is usually approximated as a sum of two-body interactions for each element pair.  As such, it has been identified as a prime target to improve the overall performance of the method.

Despite their apparent simplicity, the two-body repulsive potentials commonly used in DFTB are often sufficiently accurate, as DFTB is widely used in the community to study structural, electronic, optical and vibrational properties of varied physical systems. 
Even if any system can in theory be used to fit and generate such bond-based repulsive potentials\cite{oliveira2015dftb}, choosing the reference is nevertheless challenging. Identifying and isolating the target energy of one specific bond requires the   consideration of suitable systems. Isolated objects are a first intuitive candidate, as one easily obtains data points for an energy versus distance curve by modifying the length of a bond within a molecule. By considering highly symmetric periodic systems with a uniform scaling of the lattice constant, it is also possible to get the pair potential from solids by dividing the total repulsive energy by the number of nearest-neighbour bonds. In practice, more varied geometries such as disordered systems can not easily be used as reference for the pair repulsive potential. Apart from the technical difficulties in generating the two-body potentials, there are additional, more fundamental issues. While pair potentials can account for systems where the typical bond lengths are clearly distinct (for example single to triple bonds of carbon), it is challenging to represent variations of the repulsive energy for systems with similar bond lengths.

Lately, two-body potentials for DFTB have been pushed to their limits by enforcing their convexity, leading to a perfect fit for individual polymorphs of silicon\cite{bib:akak_css_hdnnp}. A one-fits-all potential remains however elusive, hence upholding the idea that two-body interactions lack the flexibility to exhibit transferability to a large set of reference systems. Another study focused more on the technical aspects of generating a repulsive pair potential in an automated fashion\cite{bib:LC_Erep}. Other works\cite{bib:DFTB_GlobalOpti} proposed a scheme to optimize pair potential parameters alongside a global minima search for specific systems. Additionally, it has been tried to push the representation to three-(or more) body interactions by means of a linear combination of Chebyshev polynomials\cite{bib:chebyshev_DFTB}. The presented results in this and a following study\cite{bib:chebyshev_DFTB_2} are quite promising but nevertheless system specific. Eventually also ML approaches have been proposed to describe the many-body environment of an atom by means of clustering approaches\cite{bib:coulmatErep}. Such a  description seems well adapted for molecules. In fact, clustering allows to switch between several potentials for the same element pair depending on the chemical state of the system. In a very recent contribution, ML is used in the form of Gaussian Process Regression to determine the repulsive potential for organic molecules \cite{panosetti2020learning}. Such an approach can lead to high accuracy for molecular systems, but might be difficult to scale up for materials simulations. 

The question being investigated in the present work is whether it is possible to determine a single, continuous and transferable repulsive potential, embedding a purely many-body description for DFTB. We aim at a potential that describes both finite and periodic systems, that is numerically efficient and as easy to use as the traditional pair potentials. 
This starts by introducing the DFTB formalism in section \ref{sec:DFTB}, with a special focus on the repulsive energy. We then detail the machine-learning methodology used in our framework in section \ref{sec:methods}. Section \ref{sec:NN-Erep} presents results for silicon in its crystalline, liquid, amorphous and clustered form in comparison to pair repulsive potentials traditionally used in DFTB. Sporadically we compare also to two purely numerical potentials for silicon trained to reproduce the total energy rather than focusing only on the repulsive energy. In fact, it has been observed that potentials obtained from ML face difficulties in treating finite and extended systems at the same time. This was traced back to long-range interactions that are difficult to describe with local characterizations of the environment and lead to the development of new strategies for ML models (see Ref.~\onlinecite{behler2021machine} and references therein). As it will be clarified later, the DFTB method with ML repulsive energies can be seen as such an approach, since long-range electrostatics are naturally accounted for. Whether this feature really leads to improved transferability will be investigated. We conclude the article with information on the numerical scaling in section \ref{sec:computational_cost} and a summary in section \ref{sec:conclusion}.

\section{\label{sec:DFTB}The DFTB formalism}
The DFTB method has been the topic of several review articles, see for example Refs.~\onlinecite{frauenheim2002asc,elstner2006scc,Elstner2014}. We will therefore discuss only the most relevant ideas in the present context. As approximation to DFT, the DFTB method is characterized by a Taylor expansion of the Kohn-Sham energy functional around a given reference density $n_0$ up to a certain order \cite{bib:dftb3}.
The most popular approach (and the one that will be used here) is labelled self-consistent-charge DFTB (SCC-DFTB) and includes density deviations with respect to the full molecular density ($n = n_0 + \delta n$) up to second order in $\delta n$. The Kohn-Sham equations resulting from the expansion read 
\begin{equation}
\label{ks}
    \sum_\nu H_{\mu\nu} c_{\nu i} = \epsilon_i \sum_\nu S_{\mu \nu} c_{\nu i}, 
\end{equation}
where $S$ denotes the overlap matrix, $c$ the molecular orbital coefficients, $\epsilon_i$ the Kohn-Sham orbital energies and $H$ is given by\cite{bib:dftb3}:
\begin{align}
\nonumber     H_{\mu\nu} &=  \langle \mu |H[n_0]| \nu \rangle \\&+ \frac{1}{2} S_{\mu \nu} \sum_C \left(\gamma_{AC} + \gamma_{BC}\right) \Delta q_C, \, \mu \in A, \nu \in B.  
\end{align}
Here $H[n_0]$ is the usual DFT Kohn-Sham Hamiltonian evaluated at the reference density $n_0$. In a basis of atomic orbitals ($\{|\mu\rangle\}$), matrix elements of $H$ can be tabulated as a function of distance between atoms, since DFTB employs a two-center approximation \cite{bib:dftb1}. The term $\gamma_{AB}$ describes the electron-electron interaction and is a function of the atomic distance ${\bf R}_{AB} = {\bf R}_A - {\bf R}_B$, while $\Delta q_A$ denotes the net Mulliken charge on atom $A$ and can be seen as a point charge representation of the density deviation mentioned above. These charges depend on the coefficients $c$ and are obtained by a self-consistent solution of Eq.~\eqref{ks}. The method is easily generalized to periodic systems \cite{bib:DFTB+1}. Solution of the Kohn-Sham equations provides access to electronic band structures, molecular orbitals and the electron density. The total energy of DFTB reads \cite{bib:dftb3}: 
\begin{equation}
\label{etot}
    E_\text{tot} = \sum_i^\text{occ} c_{\mu i}  \langle \mu |H[n_0]| \nu \rangle  c_{\nu i} + \frac{1}{2} \sum_{AB} \Delta q_A \gamma_{AB} \Delta q_B  + E_\text{rep},
\end{equation}
where the first term on the right hand side is often abbreviated as $E_\text{BS}$ (for band structure energy) and the second one as $E_\text{coul}$. 

The remaining term $E_\text{rep}$ in Eq.~\eqref{etot} is the repulsive energy and the key quantity of this article. It can be rigourously expressed as: 
\begin{equation}
\label{erep}
    \begin{split}
        E_\text{rep} = & \frac{1}{2} \sum\limits_{\substack{AB}} \frac{Z_A Z_B}{R_{AB}} - E_H [ n_0] + E_\text{xc}[ n_0 ] \\ & - \int V_\text{xc}[ n_0 ]({\bf r}) \,n_0({\bf r}) d^3r ,
    \end{split}
\end{equation}
which involves the Hartree ($E_H$) energy, as well as exchange-correlation functionals for energy ($E_{xc}$) and potential ($V_{xc}$). Atomic numbers are labelled $Z_A$. For short interatomic distances $R_{AB}$ the core repulsion dominates, while core repulsion and Hartree energy cancel exactly for large distances \cite{Foulkes1989}. This explains why this term is commonly referred to as repulsive energy. Note that Eq.~\eqref{erep} can not be exactly written as a sum of pair potentials due to the presence of the exchange-correlation functionals. 

While there have been attempts to compute $ E_\text{rep} $ directly from Eq.~\eqref{erep} within the framework of DFTB, the usual approach consists in fitting the repulsive potential to DFT total energy reference data ($E^\text{DFT}_\text{tot} $). Here one hopes to profit from the accuracy of the higher level method and remedy possible deficiencies in $E_\text{BS}$ and $E_\text{coul}$. To this end, one approximates $ E_\text{rep} $ as a sum of potentials for each atom pair:
\begin{equation}
\label{repsum}
    E_\text{rep} = \frac{1}{2} \sum\limits_{\substack{A \neq B}} V^\text{rep}_{AB} \left( R_{AB} \right),
\end{equation}
and computes 
\begin{equation}
    \label{eq:Erep_ref_def}
    E_\text{rep} = E^\text{DFT}_\text{tot} - E_\text{BS} - E_\text{coul},
\end{equation}
for varying distances between two atoms in a reference structure. This is done in such a way that the pair potentials of all other pairs either remain constant or can be assumed to have already decayed to zero. In this way one can single out and determine a specific element pair potential in the sum Eq.~\eqref{repsum}. Since the exact form of the repulsive energy depends solely on the reference density $n_0$, transferabilty is expected, such that the actual choice of reference structure should be of no importance. In practice, however, this is never truly the case and different DFTB parametrizations differ in their range of transferability. Concerning the actual functional form of the pair potential, both polynomials \cite{bib:dftb3} or splines representations \cite{frauenheim2002asc,bib:akak_css_hdnnp} have been used in the past. Common for both forms is a cutoff distance ($r_\text{cut}$), after which the potential is strictly set to zero. Too large values of $r_\text{cut}$ might hamper transferability, such that this parameter is typically set close to the nearest-neighbour distance \cite{elstner2006scc,koskinen2009density}.              

\section{\label{sec:methods}Machine learned repulsive potentials}

This section presents the conceptual tools employed (ie. Behler-Parrinello neural-networks\cite{bib:BPHDNNP}) to build flexible many-body repulsive interactions, and discusses details of the network training.

\subsection{\label{subsec:ASCF}Description of atomic environments}

The central point in the description of many-body interactions is the quantification of the chemical environment for each atom in the simulation box, leading to an atomic repulsive energy contribution. From there, we compute the total repulsive energy by adding up the atomic contributions
\begin{equation}
    E_\text{rep} = \sum\limits_{\substack{i=1}}^{N_{at}} E_\text{rep}^i.
\end{equation}

To determine $E_\text{rep}^i$, one considers a sphere of fixed radius $r_\text{cut}$ around each atom. The cutoff radius should be large enough to capture significant details of the chemical neighborhood of the atom in question, but short enough to ensure transferability. 
The atomic environment in cartesian coordinates is however not suitable as such to be fed to ML methods, as these are not invariant with respect to translation, rotation or equivalent atom exchange. Better descriptors of the atomic environment are available. 
We chose so-called Atom-Centered Symmetry Functions (ASCF)\cite{bib:ACSF}, for their frequent use in neural-networks, presented in the following section (\ref{subsec:HDNNP}). ACSF can be split into two classes. First there are radial symmetry functions, describing the environment of atom $i$ as:
\begin{equation}
    G^i_2 = \sum\limits_{\substack{j=1}}^{N_\text{neigh}^i} e^{-\eta \left( r_{ij} - r_s \right)^2} f_\text{cut} \left( r_{ij} \right),
\end{equation}
where
\begin{equation}
    f_\text{cut}(r)=
    \left\lbrace
	\begin{array}{ccc}
	  \tanh^3 \left(1 - \frac{r}{r_\text{cut}} \right)& \mbox{if} & 0<r<r_\text{cut},\\
	  0 & \mbox{else}, & \\
	\end{array}\right.
\end{equation}
is a cutoff function decaying smoothly to zero for $r=r_\text{cut}$. Three parameters $\lbrace$$\eta$, $r_s$, $r_\text{cut}$$\rbrace$ fully define a $G_2$ function: $\eta$ impacts the Gaussian part of the function, while $r_s$ shifts its center from the central atom ($r=0$). One can similarly define angular ACSF:
\begin{equation}
    \begin{split}
        G_4^i = 2^{1-\zeta} \sum\limits_{\substack{j \neq k}}^{N_\text{neigh}^i} & \left( 1 + \lambda \cos \theta_{ijk} \right)^\zeta e^{r_{ij}^2} e^{r_{ik}^2} e^{r_{jk}^2} \\ & \times f_\text{cut} \left( r_{ij} \right) f_\text{cut} \left( r_{ik} \right) f_\text{cut} \left( r_{jk} \right),
    \end{split}
\end{equation}
also depending on the angle centered on the central atom formed with two of its neighbours $\theta_{ijk}$. This class of functions has another set of parameters: $\eta$ and $r_\text{cut}$ have similar roles as for $G_2$, $\lambda = \pm 1$ shifts the minimum of the angular part to $\theta_{ijk} = 0$ or $\pi$, while $\zeta$ adjusts its width and can be seen as an angular resolution.

Each symmetry function (i.e. set of parameters) yields one real value when applied to an environment, and is completely invariant with respect to all symmetries mentioned above. Hence, one can select $N_\text{ACSF}$ sets of parameters, and compute for each atomic environment $i$ a vector $\vec{G}_i$ called descriptor, suitable to be used as input for ML methods. The exact number of ASCF isn't imposed, and one can add or remove some at will to better represent specific distances or angles. In practice, applications use between $20$ and $50$ symmetry functions ; this is often more than the amount of degrees of freedom of the system, allowing to fully grasp all meaningful information. The process of generation and selection of ASCF can be found in the literature\cite{bib:select_sf}.

\subsection{\label{subsec:HDNNP}High dimensional neural-networks}

Neural-networks (NN) enjoy a prominent spot among common ML methods. A neuron can be seen as an application that, given some input values $\vec{X}$, weights them before applying a transformation, named activation function $f^\text{act}$, to yield an output value $y$.
\begin{equation}
    y = f^\text{act} \left( b + \vec{w} \cdot \vec{X} \right).
\end{equation}
Here, weights $\vec{w}$ and the bias $b$ allow to respectively give various importance to each input value and to shift the output value. These are parameters one adjusts in order to train a network to reproduce any reference data. From now on, we will refer to a neural-network as a collection of neurons for which the input vector is independent of its output (usually called feed-forward neural-network). Furthermore, we will consider that neurons are organized in layers, the input vector of a neuron being the outputs of all neurons in the previous layer. One can then identify an input layer yielding values of the descriptor $\vec{G}_i$, an output layer providing the output of the network, and an arbitrary amount of so-called hidden layers in between. In our case, the neural-network only returns one value, the repulsive energy $E_\text{rep}^{\text{NN},i}$ of environment (or atom) $i$, since the output layer contains only one neuron. The chosen activation functions are respectively $f^\text{act}(x)=x$ for neurons in input and output layer, and $f^\text{act}(x) = \text{tanh}(x)$ for neurons located in hidden layers.

\begin{figure}[!htb]
    \centering
    \includegraphics[width=0.25\textwidth]{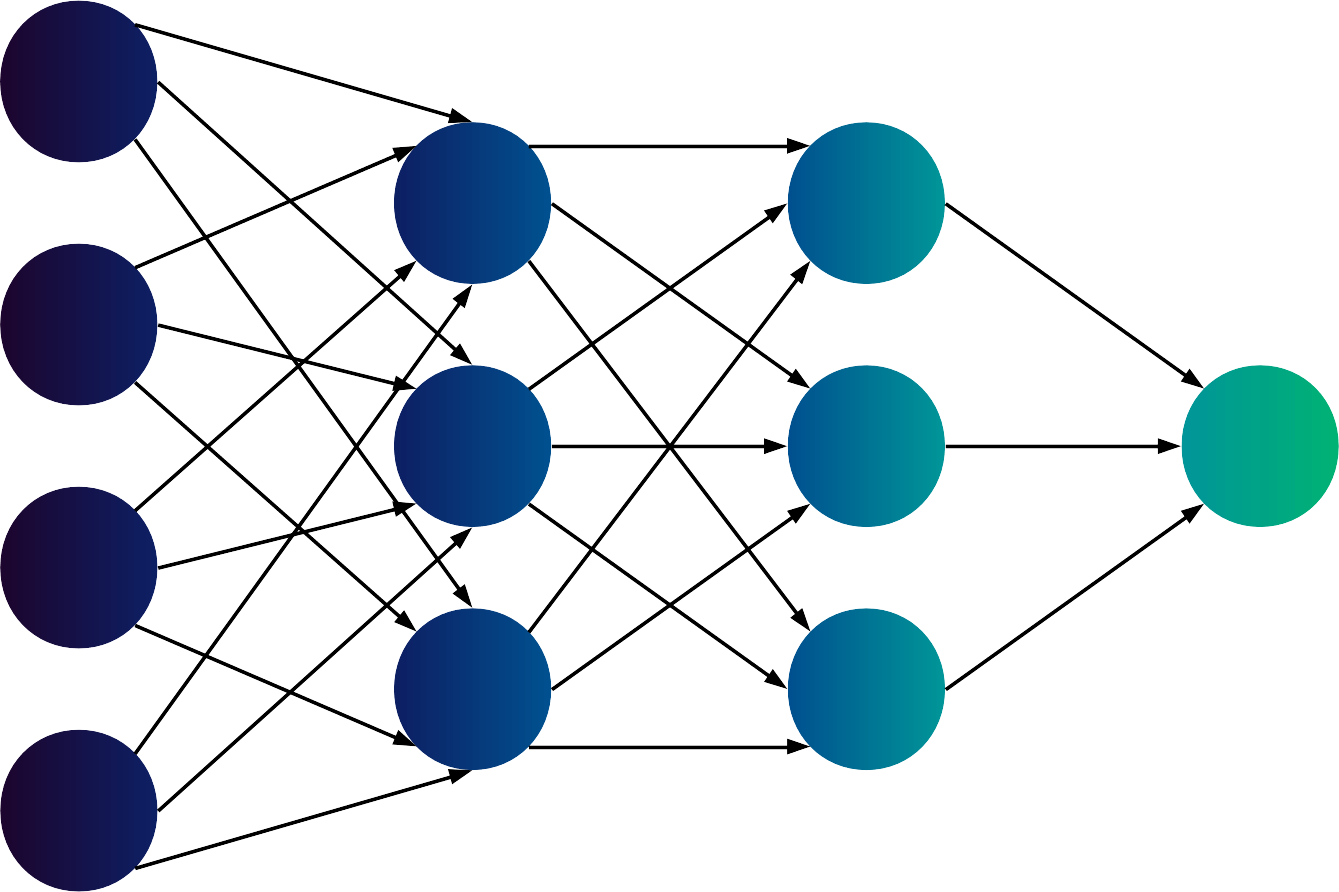}
    \caption{Structure of a $4$-$3$-$3$-$1$ feed-forward neural-network.}
    \label{fig:NN}
\end{figure}

A schematic representation of a neural-network with $4$ input neurons, $1$ output neuron and $2$ hidden layers with $3$ neurons is given in figure \ref{fig:NN}. Even for such a <<small>> network, there are $46$ parameters ($39$ weights and $7$ biases). In applications such as ours, networks can easily reach thousands of parameters, which explains the high flexibility of these tools.

\begin{figure}[!htb]
    \centering
    \includegraphics[width=0.45\textwidth]{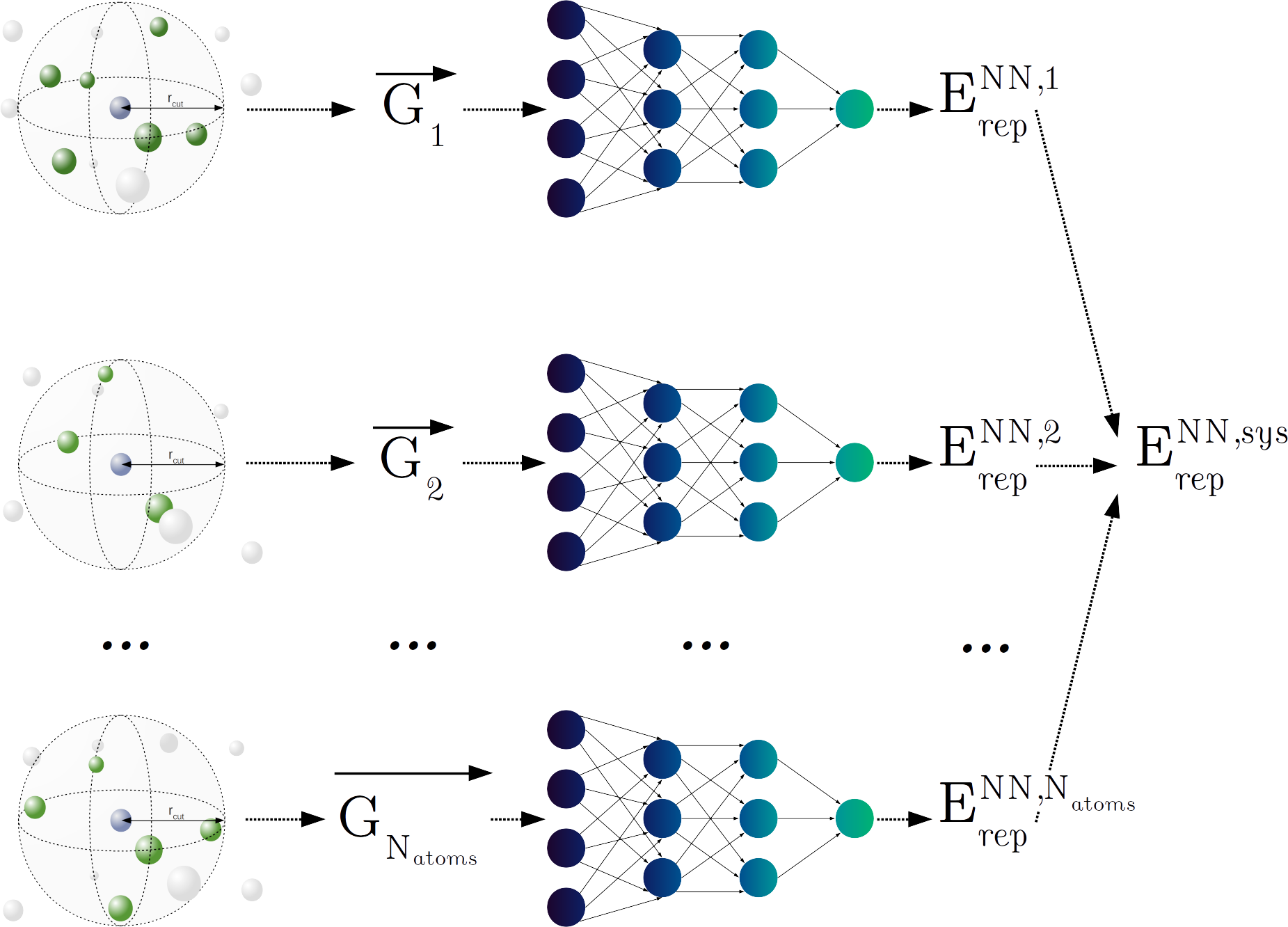}
    \caption{Concept of high-dimensional neural-network. Starting from $N_\text{atoms}$ atomic environments, the corresponding descriptors $\vec{G}_i$ are computed and fed to the neural-network to get $E^{\text{NN},i}_\text{rep}$. The sum of atomic contributions yields the total repulsive energy $E_\text{rep}^{\text{NN},\text{sys}}$, which is compared to $E_\text{rep}^{\text{ref}, \text{sys}}$ for training.}
    \label{fig:HDNNP}
\end{figure}

There is however an additional depth to account for: repulsive energies from Eq.~\eqref{eq:Erep_ref_def} (termed $E_\text{rep}^{\text{ref}, \text{sys}}$ in the following) are only given for the system as a whole and no individual atomic energies are available for training. This problem can be overcome by using $N_\text{atoms}$ copies of the neural-network, like in figure \ref{fig:HDNNP}. Training and evaluation are then done on the sum $E_\text{rep}^{\text{NN},\text{sys}} = \sum_i E_\text{rep}^{\text{NN},i}$. This additional granularity is usually termed high-dimensional neural-network\cite{bib:HDNNP}. 

Eventually, it is also possible to compute repulsive forces, i.e.~derivatives of the system repulsive energy with respect to the atomic positions.

\begin{equation}
    F_{\text{rep}, \alpha}^{\text{NN},i} = - \frac{\partial E_\text{rep}^{\text{NN},\text{sys}}}{\partial r^i_\alpha} \quad , \quad \alpha \in \left\lbrace x,y,z \right\rbrace.
\end{equation}
This can be decomposed taking the contribution of each atomic repulsive energy into account by applying the chain rule:
\begin{equation}
    F_{\text{rep}, \alpha}^{\text{NN},i} = - \sum\limits_{\substack{j=1}} \frac{\partial E_\text{rep}^{\text{NN},j}}{\partial r^i_\alpha} = - \sum\limits_{\substack{j=1}} \frac{\partial E_\text{rep}^{\text{NN},j}}{\partial \vec{G}_j} \cdot \frac{\partial \vec{G}_j}{\partial r^i_\alpha} .
\end{equation}

Before moving on to training details, let us note that the neural-network evaluation (i.e. getting $E_\text{rep}$ and $F_\text{rep}$ for a given environment) only scales with the type and number of symmetry functions and the structure of the network (see section \ref{sec:computational_cost}). Once the weights and biases have been optimized, the computation time is completely independent from the reference data. This is not the case for all other ML tools, as for instance kernel methods\cite{bib:kernel1,bib:kernel2} see their scaling directly linked to the size of the training set. This makes kernel-based methods often costly, while neural-networks show a much more appealing computation time. As the prime feature of the DFTB method is its low scaling compared to DFT (which means that typically large structures are modelled), we think neural-networks present the most suitable profile for this work.

\subsection{\label{subsec:training}Training details}
The transferability and interpolating ability of ML tools depends strongly on the quality of the training process. The task can be split into two major points: the training data, and the training algorithm. While the latter is a hot topic within the machine-learning field, it is important to note that the former may be of even greater importance. Indeed, ML methods mostly show poor extrapolation performance, and the training set has to be representative of the final application. In our case, where the objective is to develop a general purpose repulsive potential for silicon, our training set is built as follows:
\begin{itemize}
    \item Born-Oppenheimer molecular dynamics (MD) simulations of diamond (Fd$\bar{3}$m)\cite{bib:diamond1,bib:diamond2,bib:diamond3,bib:hexagonal1} in the NVT ensemble using a system consisting of $64$ atoms. One trajectory at $T=300$ K of $0.5$ ps and another one at $T=2000$ K of $0.1$ ps. 
    \item Three $1$ ps MD trajectories (respectively $T=300$ K, $T=1000$ K at normal pressure and $T=300$ K at high pressure $P$ $\in$ $\left[ 3.5, 6.7 \right]$ GPa) for other bulk polymorphs: $\beta$-$Sn$ (I$4_1$/amd)\cite{bib:betaTin1,bib:betaTin2} and the hexagonal phase (Wurtzite, P$6_3$/mmc)\cite{bib:hexagonal1,bib:hexagonal2,bib:hexagonal3} with $64$ atoms, the Weaire-Phelan cubic phase (Pm$\bar{3}$n)\cite{bib:cubic1,bib:cubic2} with $46$ atoms and monoclinic phase (C$2$/m)\cite{bib:monoclinic1} with $96$ atoms.
    \item MD at $T=2000$ K for amorphous systems. Systems contain $93$, $96$ and $100$ atoms, were obtained as described in section \ref{subsubsec:disordered} and dynamics ran for $0.2$ ps, $0.25$ ps and $0.5$ ps respectively.
    \item A $6$ ps MD at $T=3000$ K for a $128$ atoms liquid.
\end{itemize}

From those MD trajectories, we select snapshots in order to maximize the diversity of the training set. As the repulsive potential is formally defined as linking local distances and repulsive energy, this is done in two steps.
We select a first subset of approximately $10$ configurations per MD that show the most diversified distance distribution. To this, we add a second subset, where we select from the remaining configurations geometries based on their total energy per atom.
Aside from MD, we also use configurations generated in a systematic fashion:

\begin{itemize}
    \item For all five bulk systems, energy vs. volume curves, with systems containing respectively $2$ atoms ($\beta$-$Sn$ and diamond), $46$ atoms (cubic), $4$ atoms (hexagonal) and $8$ atoms (monoclinic). All lattice constants have been scaled by a factor between $0.85$ and $1.15$. For those systems, we found it was important to use only few highly compressed/stretched configurations for training. Otherwise, the high repulsive energies and forces they convey inevitably overload the cost function presented later (see eq.~\eqref{eq:cost_function}).
    \item Optimized geometries for $1000$ configurations of nano-clusters obtained in a former work\cite{bib:clusters_thomas}. The cluster sizes varied between $14$ and $77$ atoms, with $18$ to $50$ candidates for each cluster size.
    \item Compressed and elongated geometries for those same clusters.
    \item The $Si_2$ dimer with Si-Si distance within $0.80$ to $3.00$ \AA.
\end{itemize}

In total, the training set contains $2532$ systems with varied number of atoms for a total of $128523$ atomic environments.

For consistency between the results, we kept the computational parameters the same for all reference calculations. We employed {\tt VASP}\cite{bib:VASP1, bib:VASP2, bib:VASP3}, using the PBE\cite{bib:PBE} functional with PAW\cite{bib:PAW1, bib:PAW2} pseudo potentials (PAW\_PBE Si 05Jan2001 supplied with VASP). The plane wave cutoff energy was set to $400$ eV, as we found that increasing it to $520$ eV changed the energy of different bulk phases only marginally ($\lesssim 0.3$ meV/atom), while doubling the computation cost. All calculations were done spin-unpolarized.
For each system, we chose the number of K-points to reach an accuracy better than $0.3$ meV/atom for the total energy.
Eventually, we settled for a convergence threshold (\textit{EDIFF} in VASP) for the self-consistent calculations of $0.1$ meV for energies. While not optimal for some systems (EDIFF = $1$ $\mu$eV is advised for high accuracy computations), this ensures a reasonable computation time for all systems investigated.

We run DFTB single point computations for all structures mentioned above with the {\tt DFTB+}\cite{bib:DFTB+1,bib:DFTB+2} code to extract energy and forces without repulsive contributions according to Eq.~\eqref{eq:Erep_ref_def}. The electronic parametrization was taken from Ref.~\onlinecite{bib:sieck_thesis} (available as Slater-Koster set {\tt pbc-0-3} at \url{www.dftb.org}) and we nullified the repulsive potential contributions. For each system, we used the same K-points grid as in VASP to sample the reciprocal space. The maximal angular momentum was set to {\em p}. A charge convergence criterion (SCC) of $10^{-8}$ e was chosen. Simulations where self-consistency was not achieved were discarded. Similar to Eq.~\eqref{eq:Erep_ref_def} one can extract also reference values for the repulsive forces. 

For each system, we therefore have the following data
\begin{equation*}
    \left\lbrace \vec{R}_i, \vec{F}_{\text{rep}}^{\text{ref},i} \right\rbrace_{i=1}^{N_\text{atoms}}, E_\text{rep}^{\text{ref}}.
\end{equation*}

With this information, the descriptor $\vec{G}_i$ can be computed for 
the environment of each atom $i$
and the NN is trained to reproduce reference repulsive energies and forces. 
To evaluate the accuracy, we can define a cost function $\Gamma$ that is inversely proportional to the accuracy over all systems, i.e. the cost value increases for increasing error:
\begin{equation}
    \label{eq:cost_function}
    \begin{split}
    \Gamma = & \frac{1}{2 N_\text{sys}} \sum\limits_{\substack{s=1}}^{N_\text{sys}} \left[ \left( E_\text{rep}^{\text{NN},s} - E_\text{rep}^{\text{ref},s} \right)^2 + \right.\\
    & \left. \frac{\beta}{3 N_\text{atoms}^\text{s}} \sum\limits_{\substack{i=1}}^{N_\text{atoms}^s} \sum\limits_{\substack{\alpha \in \left\lbrace x,y,z \right\rbrace}} \left( F_{\text{rep},\alpha,\text{s}}^{\text{NN},i} - F_{\text{rep},\alpha,\text{s}}^{\text{ref},i} \right)^2 
    \right],
    \end{split}
\end{equation}
where $\beta=1.5$ in our application allows to adjust the relative importance between energies and forces. The next step is to minimize $\Gamma$ by choosing an optimization algorithm. We opt for the extended Kalman filter, originally\cite{bib:Kalman_original} introduced in the sixties to optimize estimators for dynamical systems. Later, Singraber et al. proposed\cite{bib:N2P2} to use it to optimize small networks (i.e. a few layers with less than a hundred neurons per layer). We found it to be roughly twice as efficient as a basic backpropagation\cite{bib:backprop} algorithm (i.e. halving reachable errors). We use the version implemented in the N2P2 software\cite{bib:N2P2_DOI}.

After extended tests, we found a $31$-$20$-$20$-$1$ topology of the NN to provide the best results, i.e. with $31$ symmetry functions ($23$ $G_2$ and $8$ $G_4$) with $r_\text{cut} = 3$ \AA\ and two hidden layers of $20$ neurons each. In the rest of this paper, we will refer to this neural-network potential as NN-E$_\text{rep}$.

\section{\label{sec:NN-Erep} Results}
In this section we investigate benefits and limitations of our approach with respect to several other silicon potentials. On one hand, we compare to the two-body potentials classically used in DFTB. In particular, we use the Splines potential of the {\tt pbc-0-3} Slater-Koster set \cite{bib:sieck_thesis}. This choice is motivated by good results of this set especially for Si clusters\cite{sieck1997structure,staab2002stability,sieck2003shape}, and because we optimized the NN with the same electronic parametrization. This eases the comparison between repulsive potentials.

On the other hand, it is also interesting to discuss the choice of learning only a part of the energy (i.e. the repulsive energy) rather than the total energy as a whole. Of course, those two methodologies are not in direct competition as they do not allow access to the same properties (details of quantum state and electronic properties), yet their comparison serves to qualitatively grasp their respective importance and the possibilities they offer. To do so, we present selected results obtained with two ML potentials from the literature used in LAMMPS\cite{bib:LAMMPS}, that try to reproduce directly DFT total energies and atomic forces. As this is not the focal point of the present work, we do not aim to extensively discuss their performance. We rather sporadically present key results that allow to contextualize and discuss the concept of machine-learned $E_\text{rep}$.

The first one is the GAP potential by Csányi and co-workers and has its parameters directly taken from Ref.~\onlinecite{bib:bartok_GAP}. It uses kernels rather than neural-networks as an interpolation tool, and relies on the Gaussian Approximation Potential (GAP) methodology\cite{bib:kernel1}. The global idea is to compare any new configuration to all known ones from the training data set. In fact, comparisons do not directly use configurations, but their descriptors. The employed descriptors are based on the overlap of atomic densities\cite{bib:on_representing}. GAP is intended to be a general purpose interatomic potential for pure silicon, and is trained on a wide set of reference data: isolated atoms, monocrystals (diamond, $\beta$-$Sn$, simple hexagonal, diamond hexagonal, centered cubic phases bcc and bc8, fcc, compact hexagonal, tetrahedral st12), liquids, amorphous, diamond surfaces, diamond (di-)vancancies and interstitials, cracks and structures with $\text{sp/sp}^2$ hybridizations. Reference computations in Ref.~~\onlinecite{bib:bartok_GAP} have been done using CASTEP\cite{bib:CASTEP} with similar DFT parameters than our VASP computations, except for using PW91\cite{bib:PW91_1,bib:PW91_2} as the exchange-correlation functional and $250$ eV for the plane wave cutoff. The cutoff radius used by the authors is $r_\text{cut}^\text{GAP}=5$ \AA. Kernel methods require relatively large computation times. Still, this potential is known for its good reproduction of elastic, vibrational and thermal properties for pure silicon as well as defect formation energies and angular/radial distributions of disordered media.

The second one, named NNP in following, is similar to the approach in the present article and uses also high-dimensional neural-networks and Atom Centered Symmetry Functions. Training set, neural-network shape and symmetry functions ($r_\text{cut}^\text{NNP}=5.5$ \AA) are taken from Ref.~\onlinecite{bib:ML_performance}. We repeated the training process, arriving at accuracies similar to the ones presented in the original article (RMSE $\sim 7$ meV/atom and $\sim 0.12$ eV/\AA). In this case the training set contains: bulk crystals (body centered cubic, face centered cubic and diamond) with gradually increasing strain, molecular dynamics on these systems at various temperatures (including simulations above the melting temperature) with or without vacancies, and surfaces. Details are given in the original publication\cite{bib:ML_performance}. Reference computations were done with VASP using the same parameters as in the present study, except for a $520$ eV cutoff of plane waves. This potential was tested mainly on structural and energetic properties of crystals.

The VASP parameters in all following computations to evaluate the accuracy of each method are the same as used to train the neural-network (see parameters in section \ref{subsec:training}). 

\subsection{\label{subsec:periodic_systems}Bulk systems}

This part is grouped in two sections. First, we discuss various properties for Si crystals before focusing on disordered systems.

\subsubsection{\label{subsubsec:bulks}Crystal polymorphs}

\begin{figure}[!htb]
    \centering
    \begin{minipage}[b]{0.95\linewidth}
      \centering
      \includegraphics[width=\textwidth]{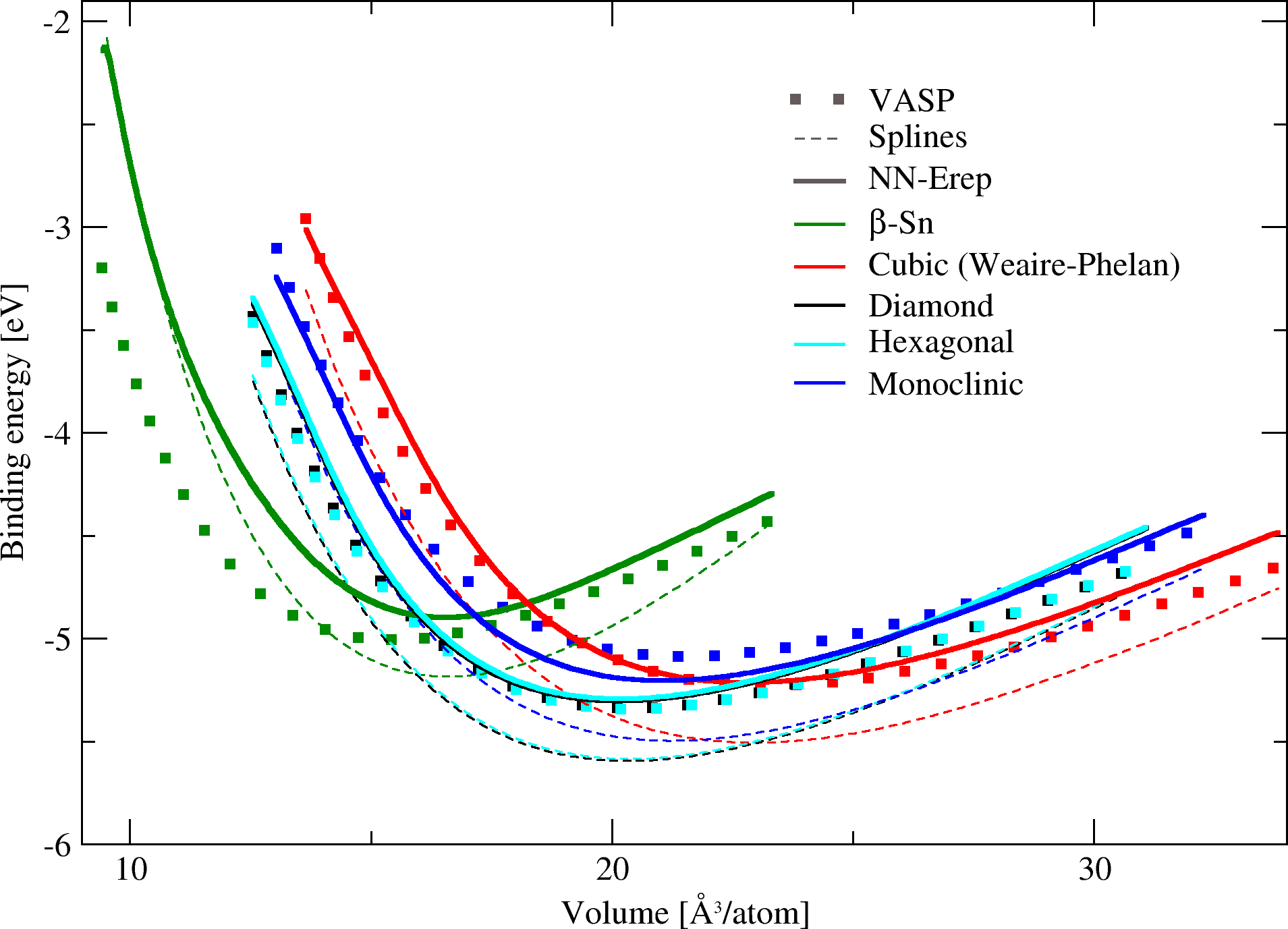}
      {(a) DFT (VASP) versus DFTB (Splines and NN-E$_\text{rep}$)}
      \vspace{2ex}
    \end{minipage} \\
    \begin{minipage}[b]{0.95\linewidth}
      \centering
      \includegraphics[width=\textwidth]{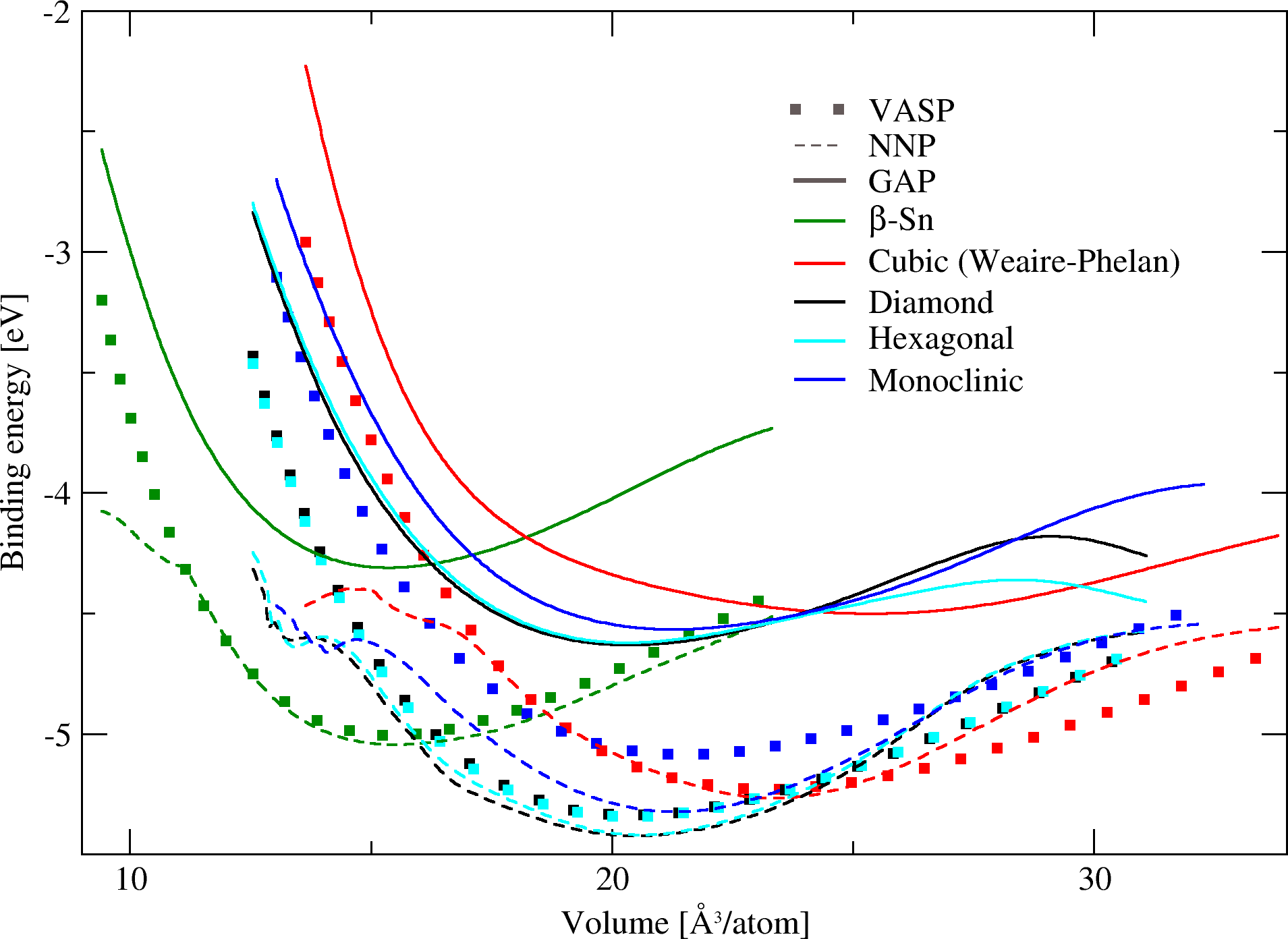}
      {(b) DFT (VASP) versus full-ML potentials (NNP and GAP)}
      \vspace{2ex}
    \end{minipage}
    \caption{Comparison of energy vs. volume curves for monocrystals computed with DFT/PBE (VASP) and several Si potentials.}
    \label{fig:evsv_bulks}
\end{figure}

The relative stability of different polymorphs serves as a first benchmark. We compute the cohesive energy according to 
\begin{equation}
    E_b = E_\text{tot} - N_\text{atoms} \times E_\text{iso},
\end{equation}
where $E_\text{tot}$ is the total energy of the unit cell with $N_\text{atoms}$ atoms and $E_\text{iso}$ is the energy of an isolated atom. We provide the evolution of $E_b$ with respect to the volume of the unit cell in figure \ref{fig:evsv_bulks}. Primitive cells contain $2$ atoms for $\beta$-$Sn$ (I$4_1$/amd) and diamond (Fd$\bar{3}$m), $46$ atoms for cubic (Weaire-Phelan - Pm$\bar{3}$n), $4$ atoms for hexagonal (P$6_3$/mmc) and $8$ atoms for monoclinic (C$2$/m) Si, respectively. The volume of the unit cell is changed by scaling lattice parameters and atomic positions by a given value. The initial cells and positions were optimized with VASP.

\begin{figure*}[!htb]
  \centering
  \begin{minipage}[b]{0.3\linewidth}
    \centering
    \includegraphics[width=\textwidth]{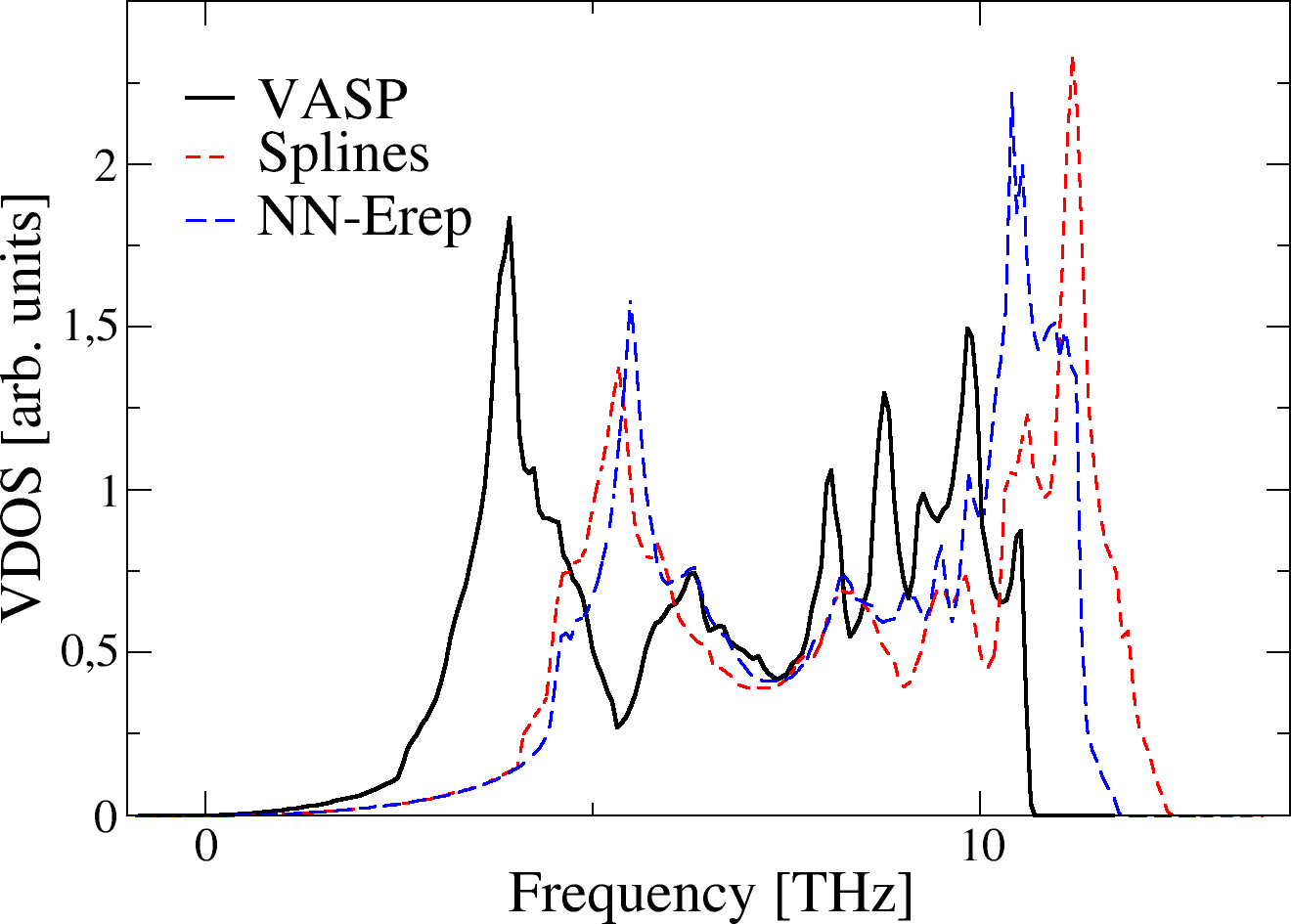}\\ \vspace{2ex}
    {(a) $\beta$-$Sn$} 
    \vspace{2ex}
  \end{minipage}
  \hfill
  \begin{minipage}[b]{0.3\linewidth}
    \centering
    \includegraphics[width=\textwidth]{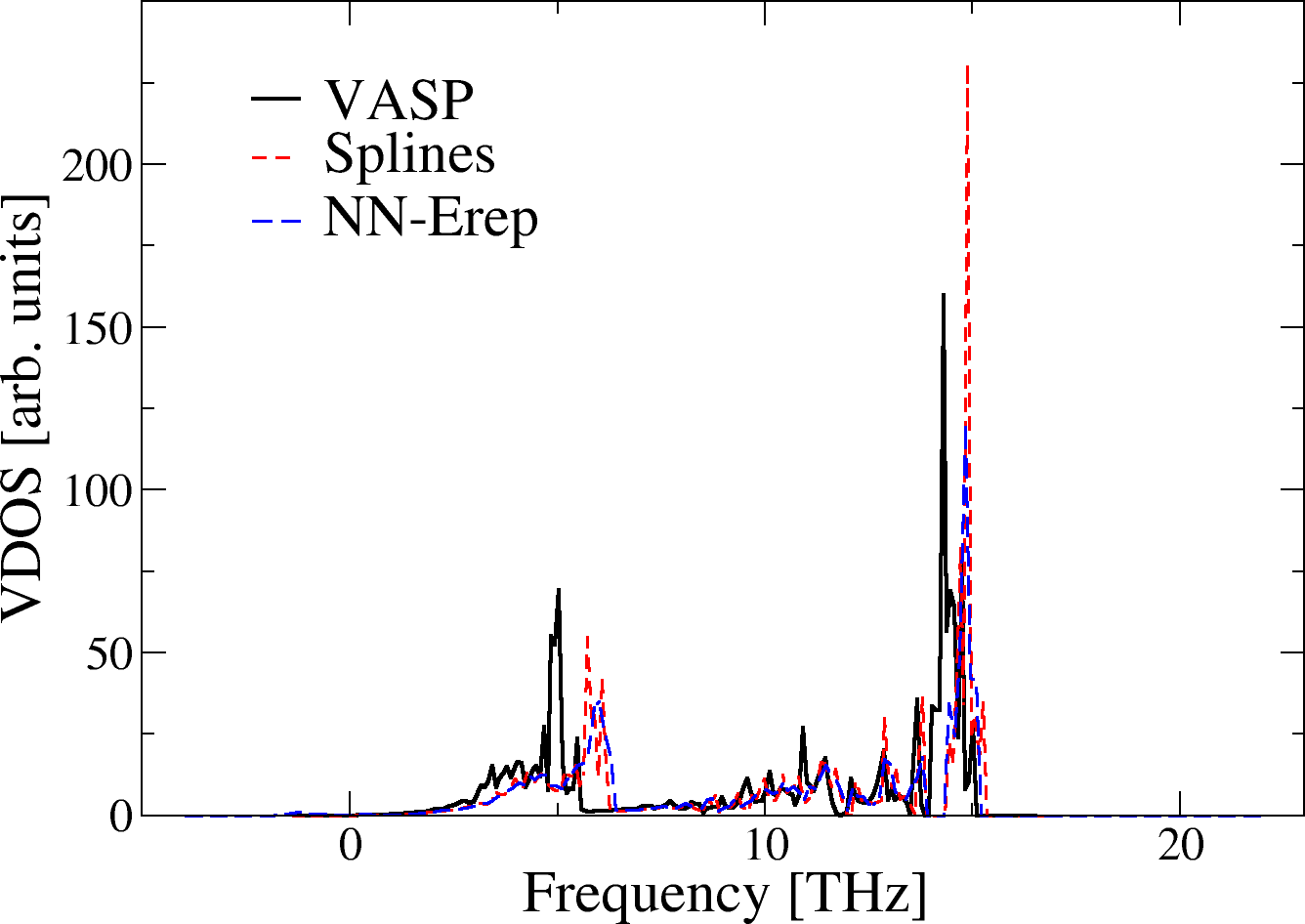}\\ \vspace{2ex}
    {(b) Cubic (Weaire-Phelan)}
    \vspace{2ex}
  \end{minipage} 
  \hfill
  \begin{minipage}[b]{0.3\linewidth}
    \centering
    \includegraphics[width=\textwidth]{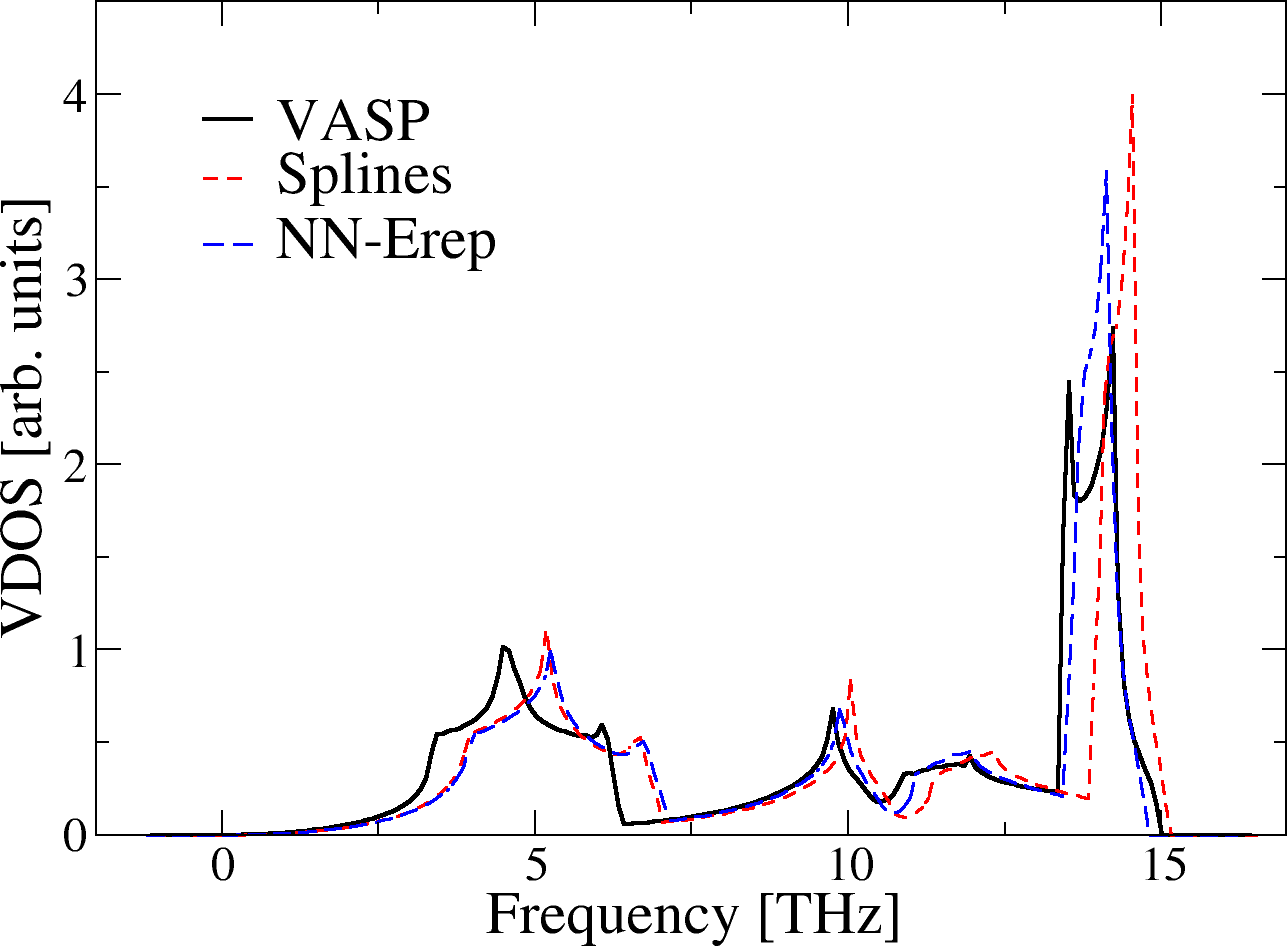}\\ \vspace{2ex}
    {(c) Diamond} 
    \vspace{2ex}
  \end{minipage} \\
  \begin{minipage}[b]{0.3\linewidth}
    \centering
    \includegraphics[width=\textwidth]{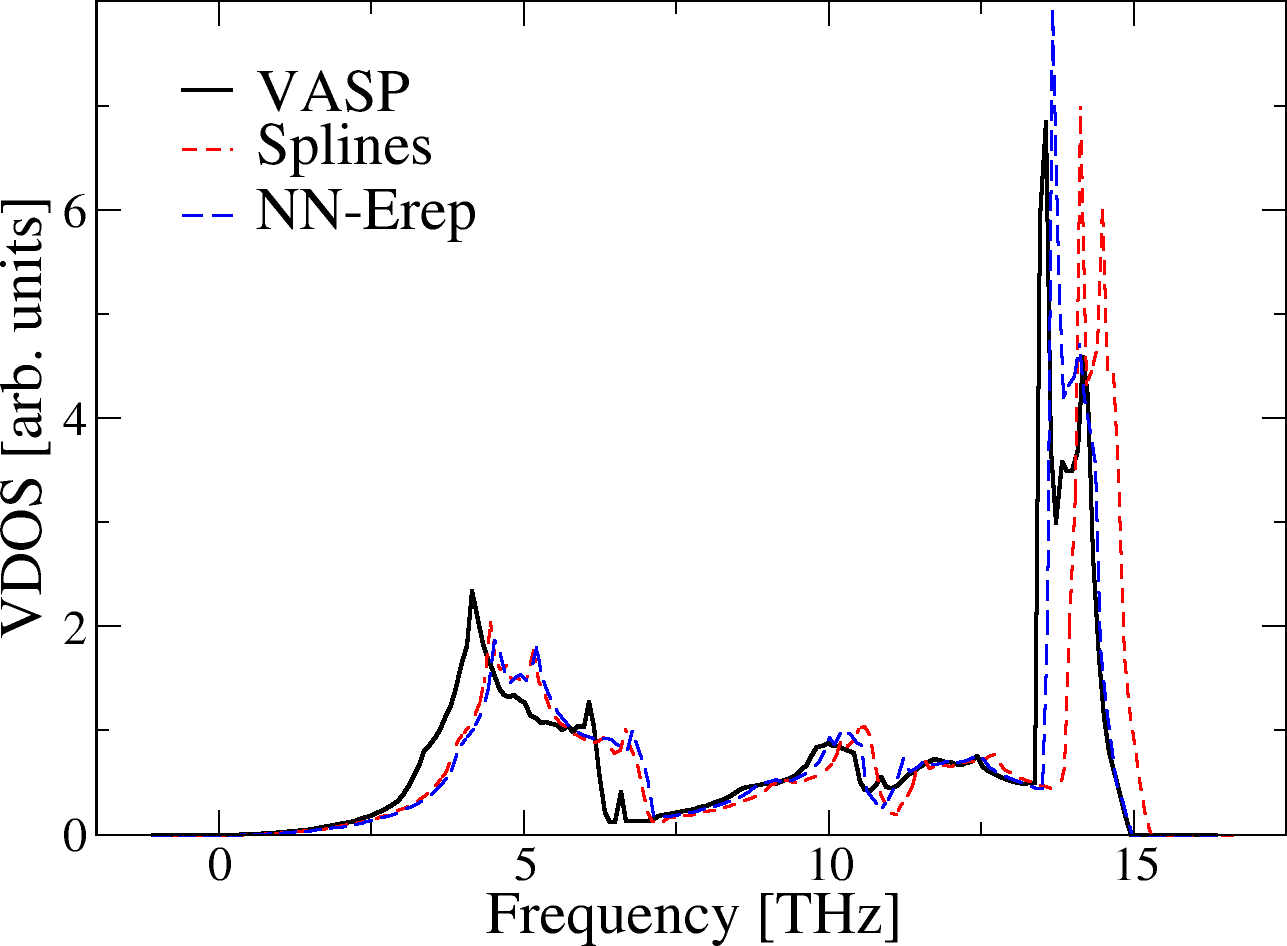} \\ \vspace{2ex}
    {(d) Hexagonal}  
  \end{minipage} 
  \hspace{0.05\textwidth}
  \begin{minipage}[b]{0.3\linewidth}
    \centering
    \includegraphics[width=\textwidth]{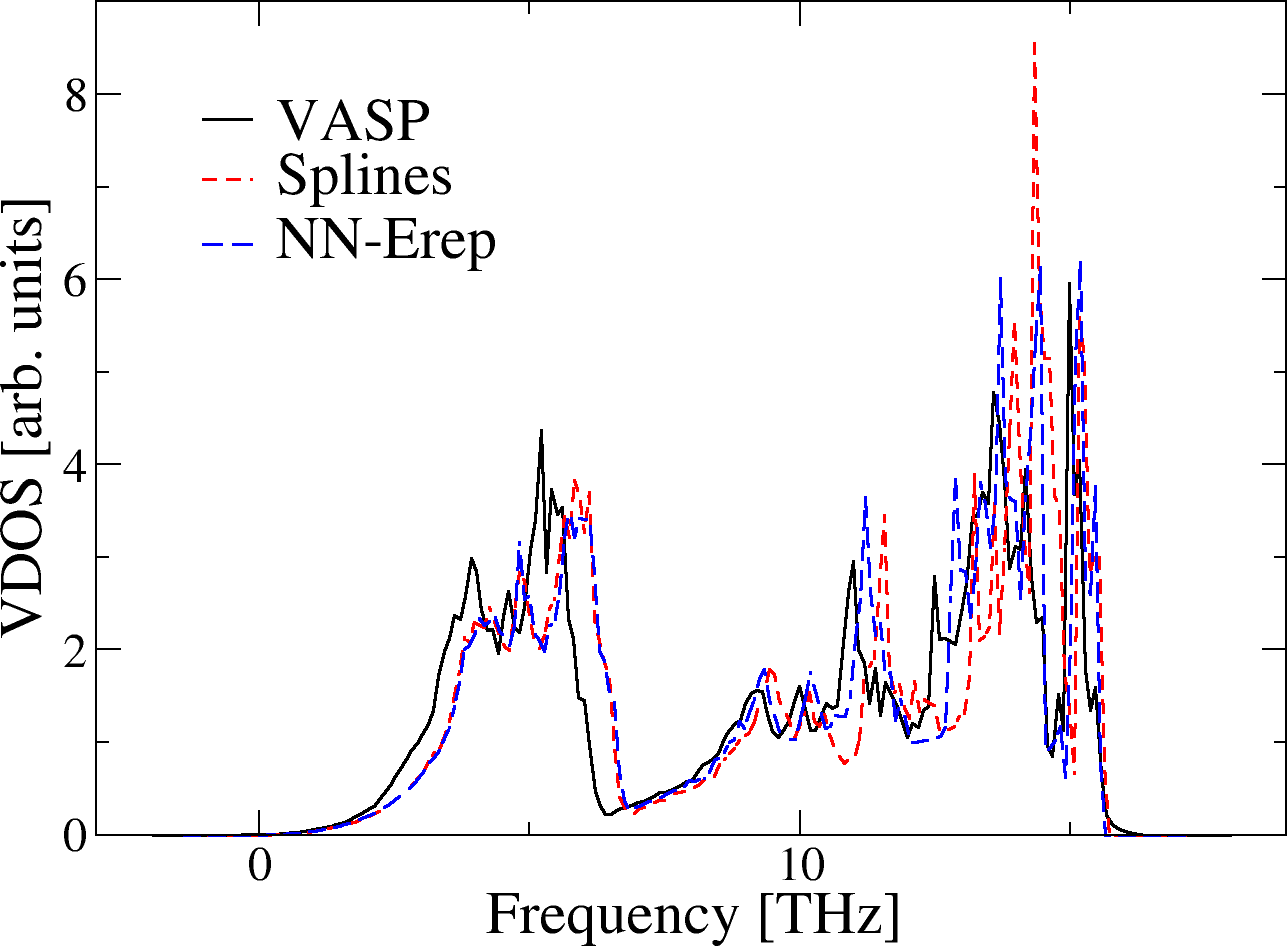} \\ \vspace{2ex}
    {(e) Monoclinic} 
  \end{minipage}
  \caption{\label{fig:VDOS_bulks_DFTB} Comparison of vibrational density of states for pure silicon bulks polymorphs computed with VASP and DFTB with different repulsive potentials.}
\end{figure*}

Overall, NN-E$_\text{rep}$ yields better results than Splines for all system but $\beta$-$Sn$. Here we observe a similar repulsion energy at small volumes for both methods. This is due to the DFTB energy contributions $E_\text{BS}$ and $E_\text{coul}$ that dominate for small volumes of $\beta$-$Sn$. In other words, the NN is only able to bring a sensitive correction for larger volumes. These results confirm recent efforts\cite{bib:akak_css_hdnnp} showing that the additional flexibility of many-body repulsive interactions lead to higher transferability for cohesive energies compared to pair potentials.

For large volumes, associated with long interatomic distances, full-ML potentials (GAP and NNP) lack reference data and show spurious behaviour with slightly decaying energies. In the case where only $E_\text{rep}$ is fitted, those large distances are handled by electronic contributions to the DFTB energy. Purely numerical approaches lack this robustness, while NN-E$_\text{rep}$ has a smaller complexity to handle. Moreover, in the case of NNP very short distances are poorly represented, with unphysical minima appearing as the training set lacks reference points. 

Energy vs. volume curves show a rather good shape for bulks with NN-E$_\text{rep}$, seemingly pointing to an accurate force prediction. To assert that statement, we focus on the prediction of vibrational properties, as these directly result from the interatomic forces. Starting from DFT/VASP optimized cells and positions, we optimize the atomic positions with both repulsive potentials. From there, we use the phonopy\cite{bib:phonopy} code to replicate cells, yielding $250$ atoms for $\beta$-$Sn$, $368$ atoms for cubic, $128$ for diamond, $256$ for hexagonal and $216$ atoms for monoclinic Si. We obtain the vibrational density of states (VDOS) using a $30\times30\times30$ Monkhorst-Pack\cite{bib:monkhorst_pack} grid to sample the Brillouin zone. The VDOS for all five bulk systems obtained with DFT/VASP and DFTB (Splines and NN-E$_\text{rep}$) is given in figure \ref{fig:VDOS_bulks_DFTB}.

For all five polymorphs investigated, one first notices that the VDOS at lower frequencies look rather insensitive to the repulsive potential used. Hence, one can assume that the acoustic modes are mostly ruled by the longer range electronic terms of DFTB ($E_\text{BS}$ and $E_\text{coul}$). For cubic (\ref{fig:VDOS_bulks_DFTB}.b) and monoclinic (\ref{fig:VDOS_bulks_DFTB}.e) Si, both repulsive potentials are really similar. The improvement brought by the present approach NN-E$_\text{rep}$ is more noticeable on optical modes of $\beta$-$Sn$ (\ref{fig:VDOS_bulks_DFTB}.a), without suceeding to be truly representative. Eventually, while Splines predict a decent VDOS for diamond (\ref{fig:VDOS_bulks_DFTB}.c) and hexagonal (\ref{fig:VDOS_bulks_DFTB}.d) Si, the neural-network repulsive potential is more accurate, reproducing the VDOS almost perfectly for intermediate frequencies and optical modes ($\omega \geq 7.5$ THz). 

The ability of NN-E$_\text{rep}$ to outperform Splines on properties of diamond is particularly interesting, as both repulsive potentials have been trained on this polymorph (the {\tt pbc-0-3} Splines have been fitted to energy-distance data of diamond and simple cubic Si). Hence, it seems that many-body potentials are simultaneously able to be more accurate on specific systems while also showing a greater capability to handle the complexity carried by a varied training set.

\begin{figure}[!htb]
  \centering
  \begin{minipage}[b]{0.6\linewidth}
    \centering
    \includegraphics[width=\textwidth]{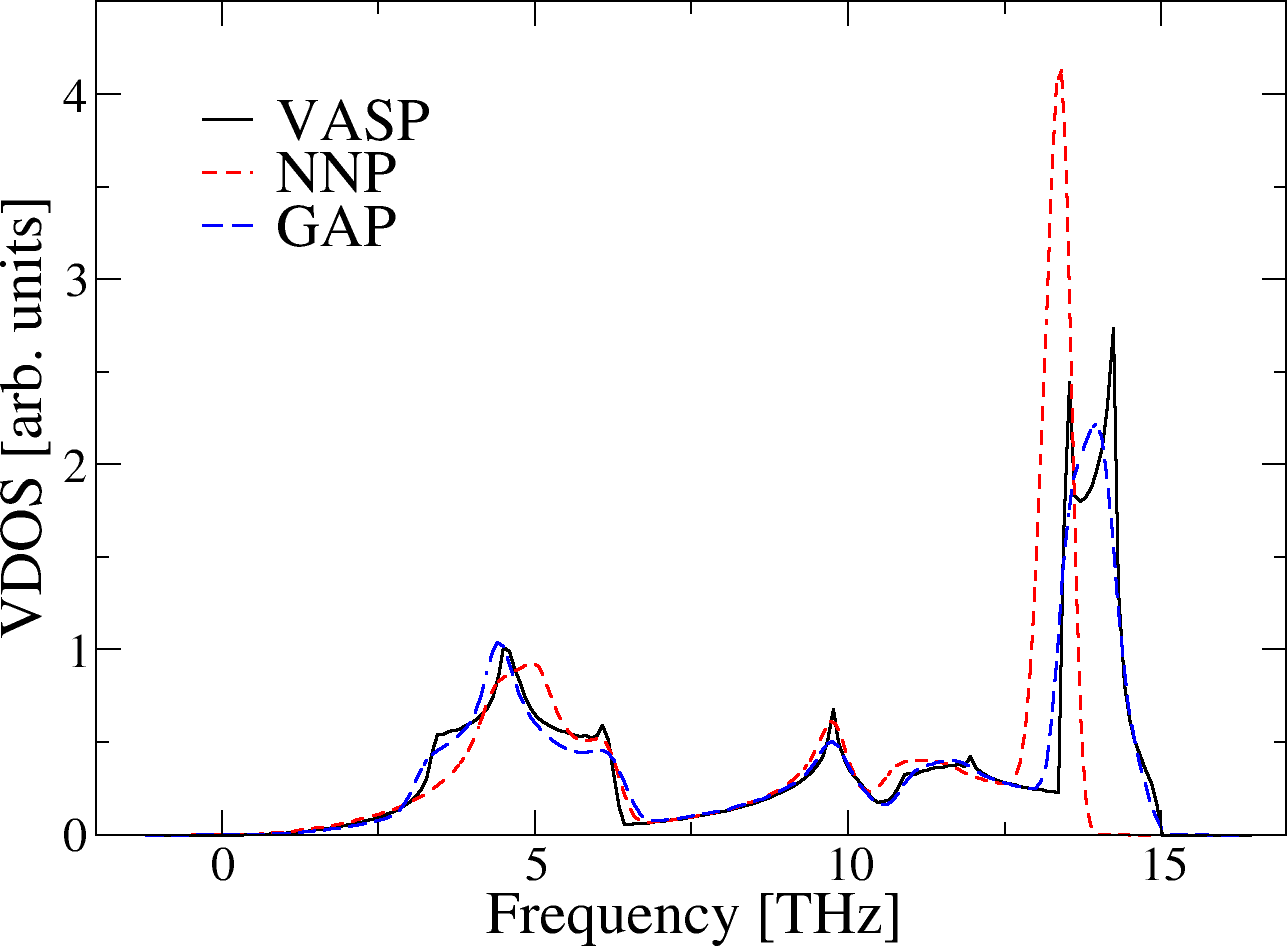}\\ \vspace{2ex}
    {(a) Diamond.} 
    \vspace{2ex}
  \end{minipage} \\
  \begin{minipage}[b]{0.6\linewidth}
    \centering
    \includegraphics[width=\textwidth]{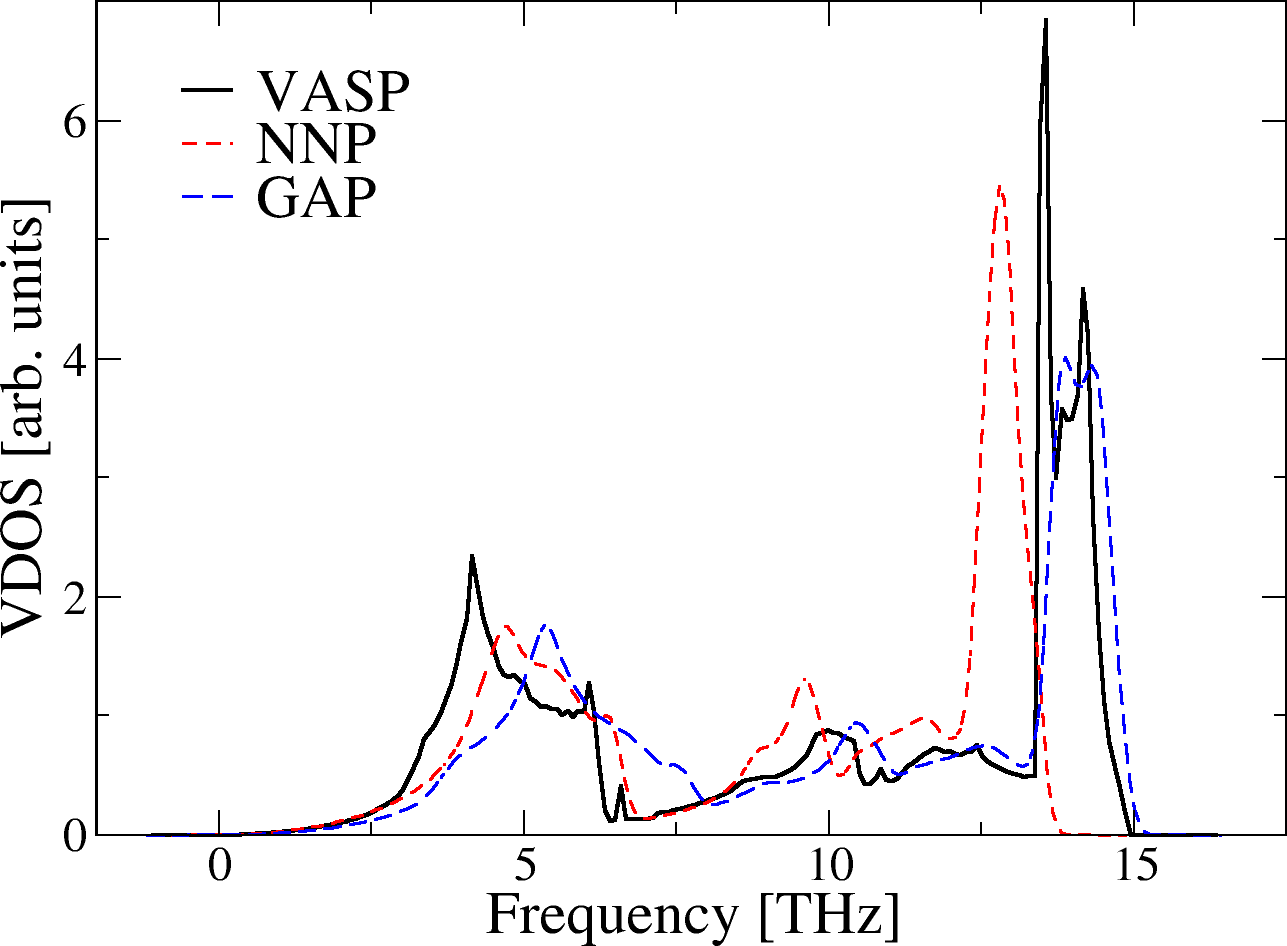}\\ \vspace{2ex}
    {(b) Hexagonal.}
    \vspace{2ex}
  \end{minipage}
  \caption{\label{fig:VDOS_LMP}Vibrational density of states for some Si polymorphs predicted by two fully-numerical approaches using machine-learning tools.}
\end{figure}

We report in figure \ref{fig:VDOS_LMP} the vibrational densities of states of diamond and hexagonal silicon obtained using GAP and NNP. Those have been computed using both potentials in LAMMPS\cite{bib:LAMMPS} through the phonoLAMMPS interface with the same parameters previously used. We only report the VDOS for these two crystals to discuss some key points.

For diamond with the GAP potential (figure \ref{fig:VDOS_LMP}.a), we observe a perfect fit to the DFT results for all frequencies. Here, GAP is not hampered by any external contribution, as opposed to our approach NN-E$_\text{rep}$ (figure \ref{fig:VDOS_bulks_DFTB}.c) that can not  fully correct the low frequency acoustic modes dominated by electronic contributions to the DFTB energy. However, this unraveled flexibility of purely numerical tools is not always linked to a better representation: for instance in figure \ref{fig:VDOS_LMP}.b, neither GAP nor NNP outperform NN-E$_\text{rep}$ on hexagonal VDOS.

For systems close to the training data, the predicted energies (figure \ref{fig:evsv_bulks}) and forces (figure \ref{fig:VDOS_bulks_DFTB}) shows significant improvements with respect to Splines. It is interesting to investigate how the neural network NN-E$_\text{rep}$ performs on systems not directly known. To do so, we study the stability of several point defects for diamond silicon: the diamond vacancy, and two different interstitials (tetrahedral and hexagonal). We start from an ideal diamond unit cell, with atomic positions and lattice constants optimized by VASP, containing a large number of atoms. As in previous studies\cite{bib:bartok_GAP, bib:Si_defects} for similar applications, we choose $N_\text{atoms} = 216$. We then introduce the desired defect, either by adding an atom in the suitable position for interstitials or by removing one atom to generate the vacancy. We then relax atomic positions with each investigated method (DFT/VASP, DFTB+Splines and DFTB+NN-E$_\text{rep}$). 

\begin{figure}[!htb]
    \centering
    \includegraphics[width=0.475\textwidth]{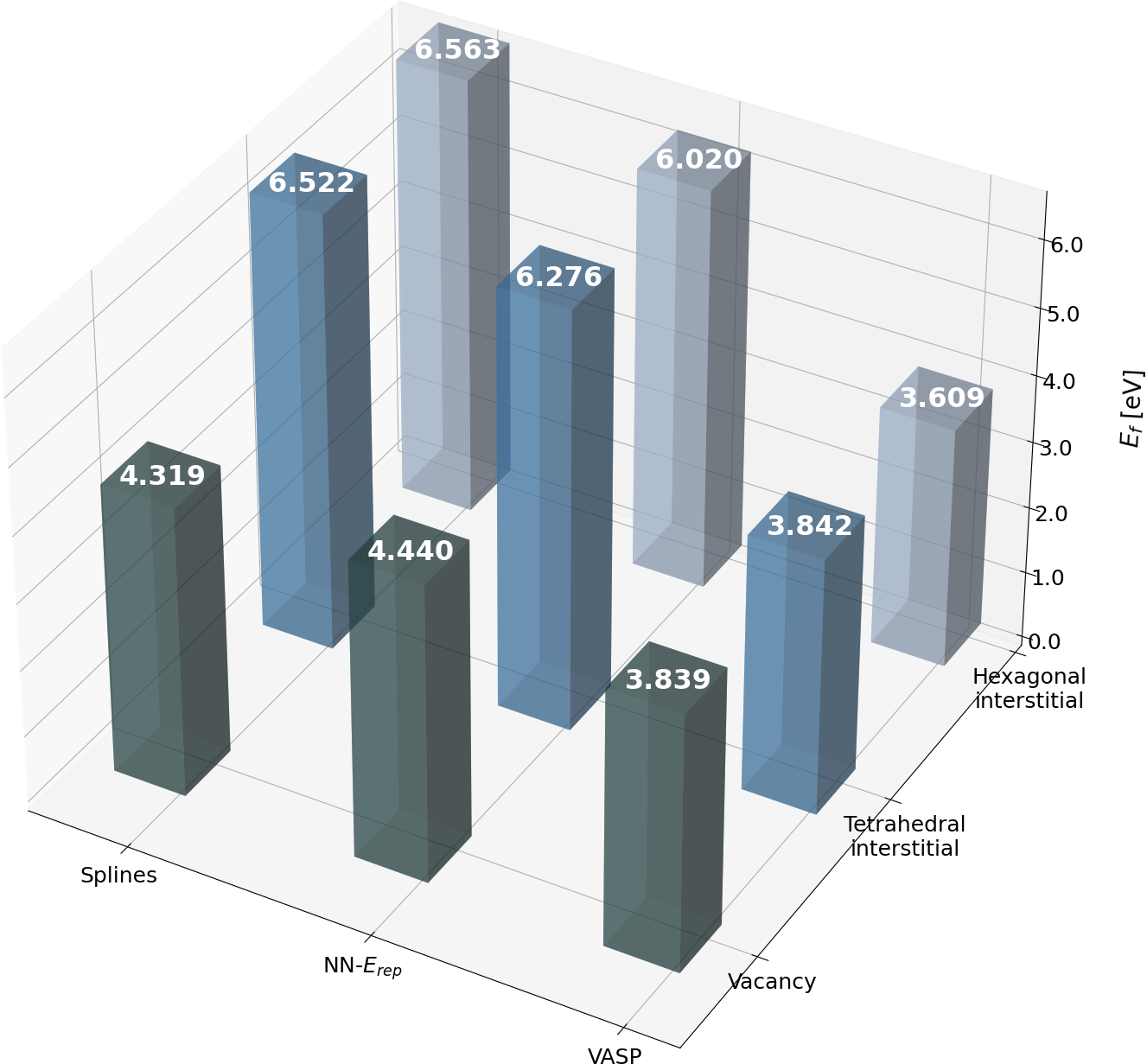}
    \caption{Point defect formation energy computed with DFT/VASP and DFTB with different repulsive potentials.}
    \label{fig:defect_formation_energy}
\end{figure}

The defect formation energy $E_f$ is then computed as 
\begin{equation}
    E_f = E_{\text{tot}}^\text{defect} - N_\text{atoms}^\text{defect} \times E_{\text{tot,at}}^{\text{diamond}},
\end{equation}
where $N_\text{atoms}^\text{defect} = N_\text{atoms} \pm 1$ denotes the number of atoms for the vacancy or interstitial cell, $ E_{\text{tot}}^\text{defect} $ the corresponding total energy and $E_{\text{tot,at}}^{\text{diamond}}$ is the energy per atom in a perfect diamond system. We report the computed formation energy for each defect and each method in figure \ref{fig:defect_formation_energy}.

The first striking tendency we observe is that DFTB underestimates the stability of defects with respect to DFT/VASP. The impact of the repulsive potential is marginal, and DFTB seems to penalize even more interstitials. Hence, vacancies are always predicted as the most stable defect, in disagreement with the DFT reference. However, the many-body repulsive potential NN-E$_\text{rep}$ correctly predicts the hexagonal interstitial as the most stable of the two, with almost the same energy difference as predicted by DFT/VASP (VASP: $\Delta E_f \approx 0.233$ eV, NN-E$_\text{rep}$: $\Delta E_f \approx 0.256$ eV). On the other hand, Splines seems to amalgam both interstitials and predicts rather similar formation energies. In this regard, it seems that the neural network is able to discriminate between several point defects, but the correction is not sufficient to fully compensate the shortcomings of DFTB.

The potential NN-E$_\text{rep}$ has not been explicitly trained on any point defect configuration. The slight performance gain can be explained by the inclusion of disordered systems in the training data. For such systems, some local configurations contain under(over)-coordinated atoms and are indeed similar to defects configurations in diamond silicon. This situation cannot be qualified as extrapolation \textit{per se}, but is a clear case of transferability to untargeted properties due to the ability of ML approaches to train on a large variety of reference data.

\subsubsection{\label{subsubsec:disordered}Disordered systems}

\begin{figure}[!htb]
    \centering
    \begin{minipage}[b]{0.95\linewidth}
        \centering
        \includegraphics[width=\textwidth]{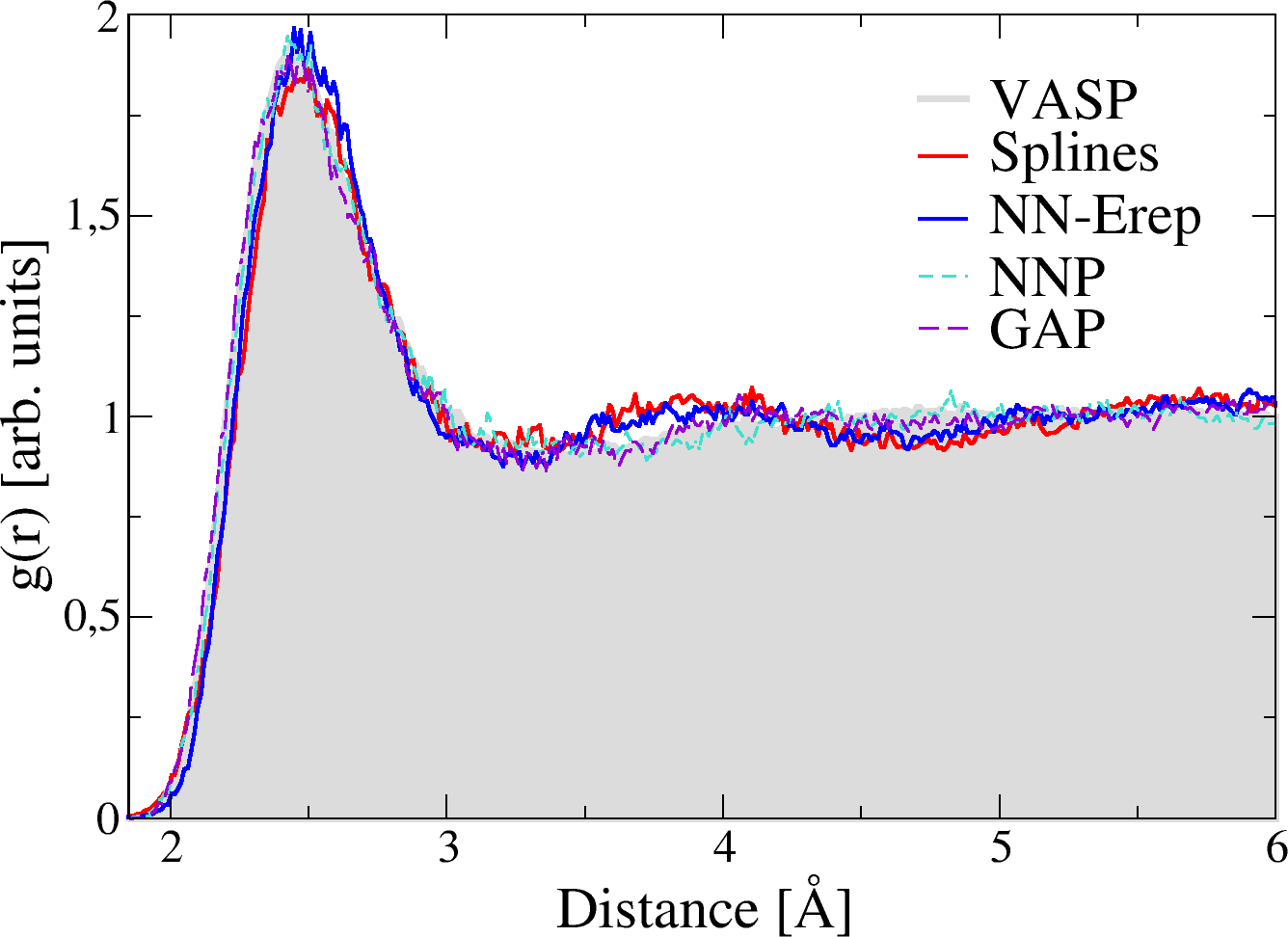}\\ \vspace{2ex}
        {(a) Radial distribution functions for liquid silicon at $T=3000$ K with neighbours less distant than $6$ \AA.} 
        \vspace{2ex}
    \end{minipage} \\
    \begin{minipage}[b]{0.95\linewidth}
        \centering
        \includegraphics[width=\textwidth]{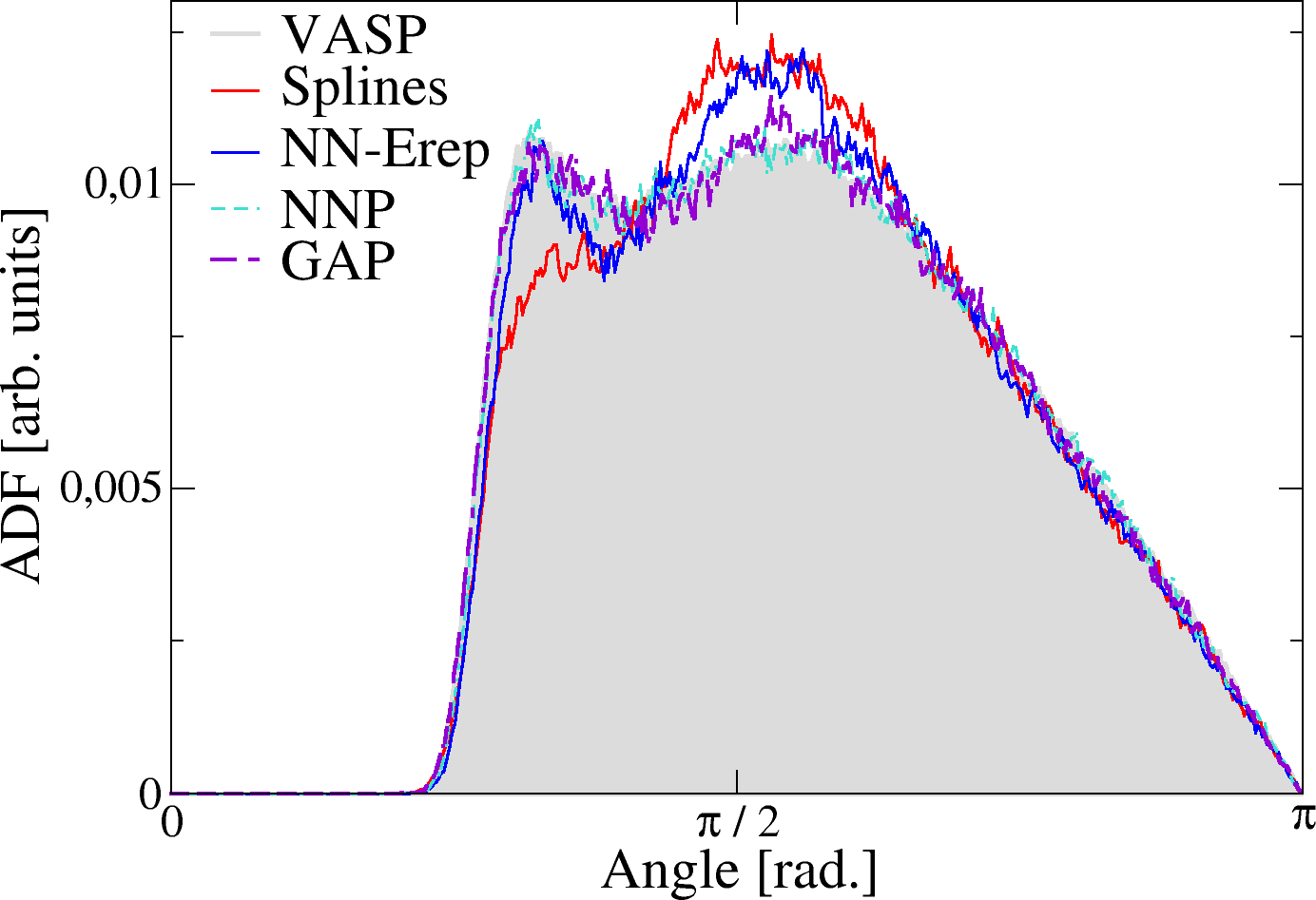}\\ \vspace{2ex}
        {(b) Angular distributions function for liquid silicon at $T=3000$ K with neighbours less distant than $3$ \AA.} 
        \vspace{2ex}
    \end{minipage}
    \caption[ Radial and angular distribution functions for liquid silicon at $T=3000$ K.]
    {\small Radial and angular distribution functions for liquid silicon at $T=3000$ K.} 
    \label{fig:liq_3000K_DFTB}
\end{figure}

Disordered systems such as liquids and amorphous systems are also of key interest, as DFTB pair repulsive potentials are typically constructed on the basis of high symmetry crystals or simple molecules. Therefore it is difficult to optimize them for unordered systems. We first focus on liquids. For such systems we evaluated the distribution of distances and/or angles within the system.
These properties give a rather good overview of the order of the system and depend critically on the temperature. The neural-network repulsive potential has been trained on temperatures up to $T=3000$ K and is expected to be representative of all liquids where $T \leq 3000$ K. To assert this, we focus on radial and angular distribution
functions extracted from MD simulations at $T=3000$ K for liquids of $128$ atoms. Such information is provided in figure \ref{fig:liq_3000K_DFTB}.

\begin{figure}[!htb]
    \centering
    \includegraphics[width=0.45\textwidth]{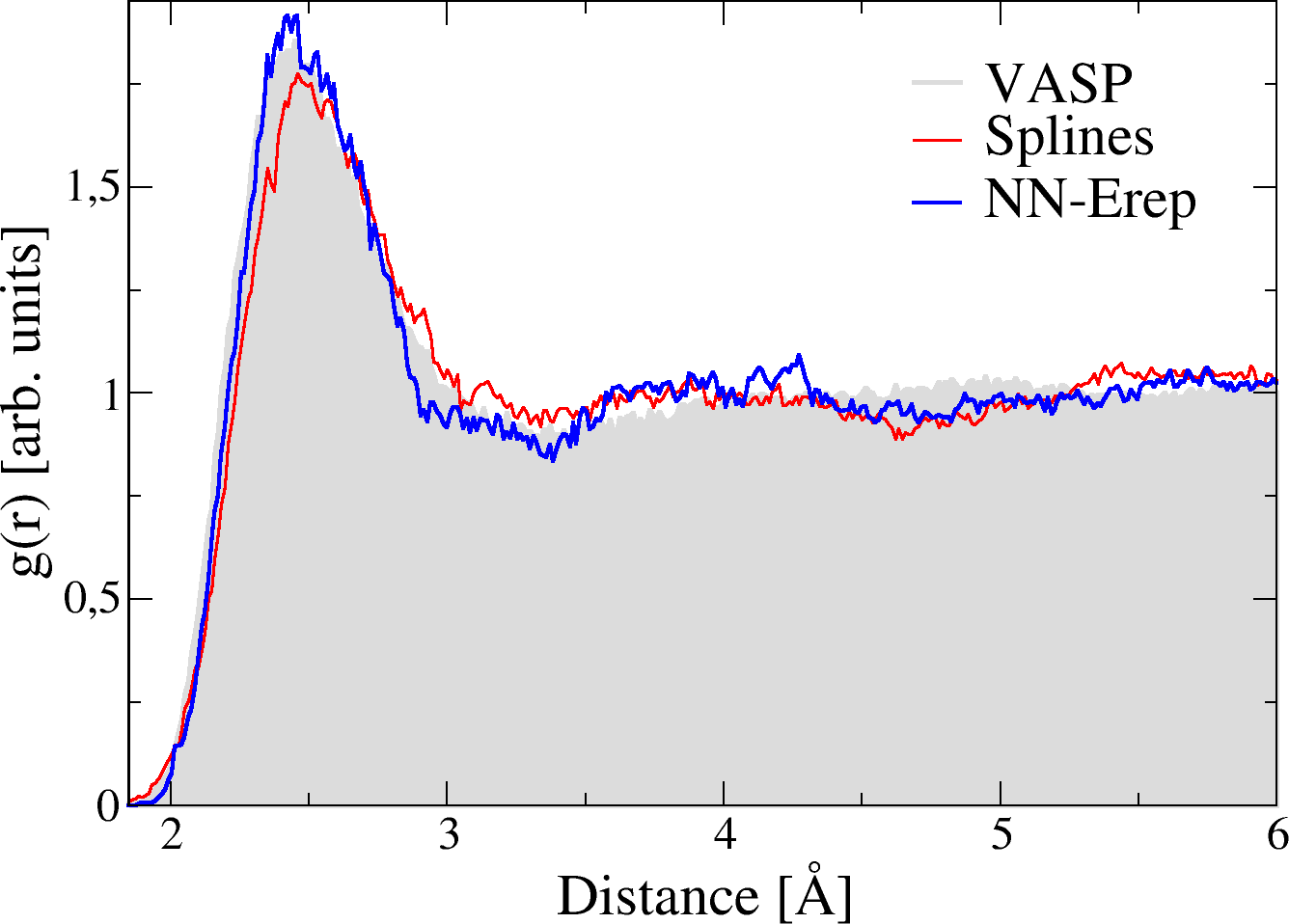}
    \caption{Radial distribution functions for liquid silicon at $T=3500$ K with neighbours less distant than $6$ \AA.}
    \label{fig:rdf_3500K_DFTB}
\end{figure}

Radial distribution functions from DFTB (regardless of the repulsive potential) over-order liquids compared to DFT/VASP and full-ML potentials (Fig. \ref{fig:liq_3000K_DFTB}.a). However, NN-E$_\text{rep}$ seems to reduce this tendency, and fits the DFT reference satisfyingly below $r_\text{cut} = 3$ \AA. As distances increase, the importance of repulsive interactions decays and the electronic terms of DFTB become the only contribution to the total energy. Their shortcomings cause most of the disagreement with DFT/VASP for $r > r_\text{cut}$. The difference between NN-E$_\text{rep}$ and Splines is more noticeable when looking at angular distribution functions (Fig. \ref{fig:liq_3000K_DFTB}.b) depending solely on the first neighbours ($r < 3$ \AA). 
Here, the Splines seem to mostly miss angles around $\pi /
3$, and highly favouring structures centered around the equilibrium angle of diamond ($\sim 109.5^\circ$). While NN-E$_\text{rep}$ also overly prefers these angles compared to DFT/VASP (and NNP/GAP), this is less marked than for Splines and allows to reproduce the first peak around $\pi / 3$. Fully numerical potentials make the most of their flexibility, and achieve a perfect representation which is not attainable when only modifying repulsive potentials. We attribute this shortcoming to the DFTB approximations of the Hamiltonian,
especially the missing of crystal field and three-center terms\cite{bib:dftb1}, which should become important in highly coordinated configurations. 
At elevated temperature higher energy regions of the potential energy surface are explored. The characteristic disorder is expected to be larger than for lower temperatures, and NN-E$_\text{rep}$ is not guaranteed to be representative. This is what we observe at $T=3500$ K (figure \ref{fig:rdf_3500K_DFTB}), where the radial distribution function shows seemingly non-physical fluctuations around $r=r_\text{cut}=3$ \AA. However, we still have $g(r)=0$ at short distances, and the potential still fulfills its repulsive role allowing no excessively close atoms.

Experimentally, the amorphous silicon structure is characterized by a density slightly inferior to the diamond crystal (a few \% lower than $2.33 \text{ g/cm}^{3}$ of diamond), the average number of bonds per atom is fewer than $4$ \cite{bib:Si_am_coord_1, bib:Si_am_coord_2} and the Si-Si-Si angle is usually centered around $108^\circ$ with a typical deviation of $\sim 10^\circ$ (see for instance Ref.~\onlinecite{bib:Si_am_angluar}).

We generated amorphous structures starting from a rapid quench ($10^{11} \text{ K/s}$) from a liquid at $3500$ K using the Tersoff potential\cite{bib:Tersoff} and a relatively large cell containing $\sim 33000$ atoms.
Smaller size periodic models with hundreds of atoms were 
extracted from the larger system and annealed at $300$ K for $10$ ps with the classical potential. Eventually, stable structures were obtained by optimizing geometries and cells with VASP, a final verification showed low pressures (of the order of few MPa).

In this way we constructed four amorphous models, three of which were used to train the neural-network repulsive potential ($N_\text{atoms} = \left\lbrace 93, 96, 100 \right\rbrace$) and one additional unseen during the training process ($N_\text{atoms} = 102$). Their respective densities are (by increasing number of atoms per unit cell) $2.31$, $2.29$, $2.28$ and $2.34 \text{ g/cm}^{3}$. Most of them exhibit only slightly lower density than the crystal reference.

Next, we evaluated the structural properties predicted by DFT/VASP and DFTB with various repulsive potentials (Splines and NN-E$_\text{rep}$ ). We relaxed atomic positions for all four amorphous models with each method, and report mean atomic coordination, mean Si-Si-Si angles and associated angular dispersions (ie. full width of the angular distribution at half maximum) in table \ref{tab:amorphous_structures}.

\begin{table}[!htb]
    \begin{ruledtabular}
    \begin{tabular}{||l||ccc||}
         & VASP & Splines & NN-E$_\text{rep}$ \\
         \hline \hline
         \textbf{93 atoms} & 4.18 neigh. & 4.10 neigh. & 4.10 neigh. \\
         Mean angle & 106.9$^\circ$ & 107.3$^\circ$ & 107.0$^\circ$ \\
         Angular dispersion & 18.1$^\circ$ & 17.1$^\circ$ & 18.0$^\circ$ \\
         \hline
         \textbf{96 atoms} & 3.99 neigh. & 3.97 neigh. & 3.97 neigh. \\
         Mean angle & 107.1$^\circ$ & 107.8$^\circ$ & 107.3$^\circ$ \\
         Angular dispersion & 16.6$^\circ$ & 15.4$^\circ$ & 15.9$^\circ$ \\
         \hline
         \textbf{100 atoms} & 3.96 neigh. & 3.96 neigh. & 3.96 neigh. \\
         Mean angle & 108.1$^\circ$ & 108.0$^\circ$ & 108.1$^\circ$ \\
         Angular dispersion & 15.8$^\circ$ & 15.8$^\circ$ & 15.2$^\circ$ \\
         \hline
         \textbf{102 atoms} & 4.00 neigh. & 3.96 neigh. & 4.02 neigh. \\
         Mean angle & 107.7$^\circ$ & 108.1$^\circ$ & 107.5$^\circ$ \\
         Angular dispersion & 16.7$^\circ$ & 16.0$^\circ$ & 17.3$^\circ$ \\
    \end{tabular}
    \caption{\label{tab:amorphous_structures} Structural properties of amorphous models predicted by DFT/VASP and DFTB with different repulsive potentials. Coordination and angular information are computed with neighbour distances up to $2.85$\AA\ - i.e. the average distance for the first minimum of all $g(r)$. }
    \end{ruledtabular}
\end{table}

With DFT/VASP the coordinations are comparable with experimental results (except for $N_\text{atoms}=93$ which is over-coordinated) and mean angles are satisfyingly close to $108^\circ$. Angular deviations are nonetheless higher than the experimental value of $10^\circ$.
We observe that all tested repulsive potentials predict coordinations in good agreement with DFT/VASP. The angular properties are slightly better for NN-E$_\text{rep}$, that yields results closer to DFT/VASP than Splines. In particular, the improvement for the angular properties is carried over to the system with $102$ atoms, even though it is not known by the neural-network. The network seems to be able to extract key local features from training amorphous data and to use it on similar systems.

\subsection{\label{subsec:isolated_systems}Nano-clusters and surface effects}

The simultaneous representation of condensed and isolated systems is a challenging issue for application using machine-learning tools (see section 5 of Ref.~\onlinecite{bib:ML_review_csanyi}). In isolated systems, quantum states are localized and slight atomic movements may yield punctual charges inducing long range electrostatic interactions. In such cases, local descriptions of the atomic environment may turn out to be too limiting. On the other hand, periodic systems tend to exhibit delocalized quantum states, with a more uniform charge distribution. This class of systems can be efficiently represented with local contributions (see for instance Refs.~\onlinecite{bib:BPHDNNP, bib:bartok_GAP, bib:ML_performance, bib:NNP_sodium}).

Some recent work by Behler et al.\cite{bib:behler_4thgen_1, bib:behler_4thgen_2} propose through their so-called 4th generation neural-networks to directly tackle this issue by including a global charge equilibration of the system to correct the sum of locally computed energies. In our case, we hope to avoid this step by simply letting the electronic contributions $E_\text{coul}$ of DFTB to handle the long-range electrostatics.

As an example of pertinent finite systems, we discuss the properties of a large set of Si clusters in the following. These configurations are taken from Ref. \onlinecite{bib:clusters_thomas} and contain cluster sizes from $14$ to $77$ atoms, with $18$ to $50$ geometries for each size. In the original article, where the intent was to determine global minima of energy per cluster size, raw configurations were obtained through the evolutionary algorithm USPEX\cite{bib:USPEX1,bib:USPEX2,bib:USPEX3} interfaced to DFTB+ to evaluate their energies. From there, the most stable geometries were optimized using the DFT code FHI-aims\cite{bib:FHI-aims} with the PBE functional. 

From these starting configurations, we reoptimized all geometries using VASP and DFTB with both repulsive potentials. The VASP global minimum is identical to the one found by Splines $28$ \% of the time, and it is among the ten most stable Spline geometries $ 81$ \% of the time. For NN-E$_\text{rep}$, those ratios are respectively $ 26$ \% and $ 79$ \%. The overall difference between both potentials is not significant, but Splines show a slightly better agreement with DFT/VASP than NN-E$_\text{rep}$. Furthermore, we observed difficulties in optimizing geometries using the neural network. Even though the cost of each optimization step is the same for both repulsive potentials (see section \ref{sec:computational_cost}), NN-E$_\text{rep}$ requires many more iterations to optimize geometries and often fails to fully converge. This is symptomatic of a noisy potential energy surface, linked to a poor reproduction of interatomic forces.

As mentioned above, fully parametric approaches are known to struggle to represent simultaneously condensed and isolated systems\cite{bib:ML_review_csanyi}. We find for each cluster size a perfect agreement with DFT/VASP of only $6$\% and $11$\% for GAP and NNP, respectively. Situations where the VASP minimum is in GAP or NNP's $10$ most stable happen respectively $50$\% and $55$\% of the time. These figures are significantly lower than the predictions of DFTB (regardless of the repulsive potential). At the same time the geometry optimization is likewise difficult with a large number of required iterations. Hence, it turns out that GAP and NNP are not especially useful to study this class of systems and we will not detail their performance any further.

\begin{figure}[!htb]
    \centering
    \includegraphics[width=0.45\textwidth]{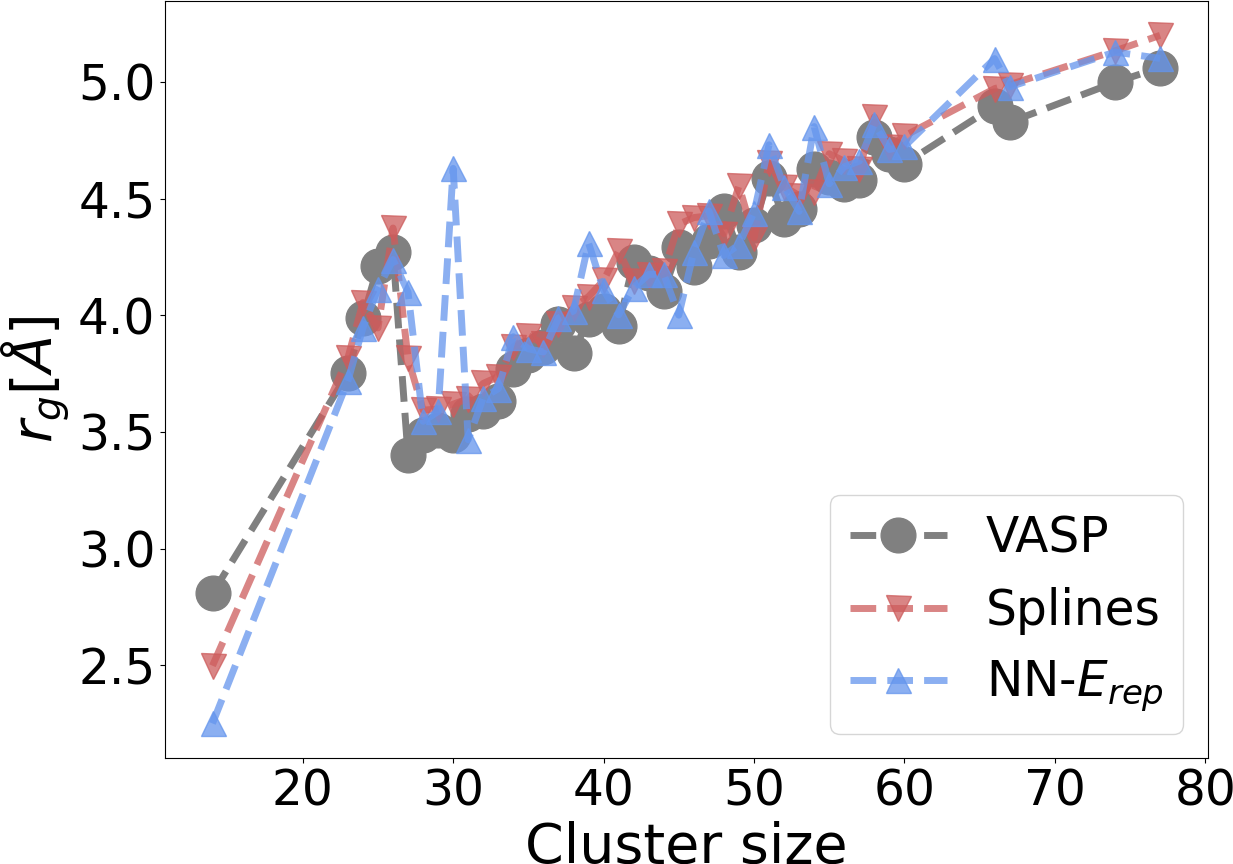}
    \caption{Gyration radii for all cluster size minima predicted by DFT/VASP and DFTB with different repulsive potentials.}
    \label{fig:gyrationRadii_DFTB}
\end{figure}

The quality of the structural predictions may be evaluated through the gyration radius of each nano-cluster:
\begin{equation}
    r_g^2 = \frac{1}{N_\text{atoms}} \sum\limits_{\substack{i=1}}^{N_\text{atoms}} \left( r_i - \Bar{r} \right)^2,
\end{equation}
which is a compactness measure. Previous studies\cite{bib:Si27_transition} on Si clusters observed a sudden drop of $r_g$ at $27$ atoms, linked to a fundamental reorganization of geometries and to a change in optical and electronic properties. The gyration radius is hence a relevant structural indicator. We report in figure \ref{fig:gyrationRadii_DFTB} the gyration radii of each predicted minimum for all cluster sizes.

\begin{figure}[!htb]
  \centering
  \begin{minipage}[b]{0.49\linewidth}
    \centering
    \includegraphics[width=\textwidth]{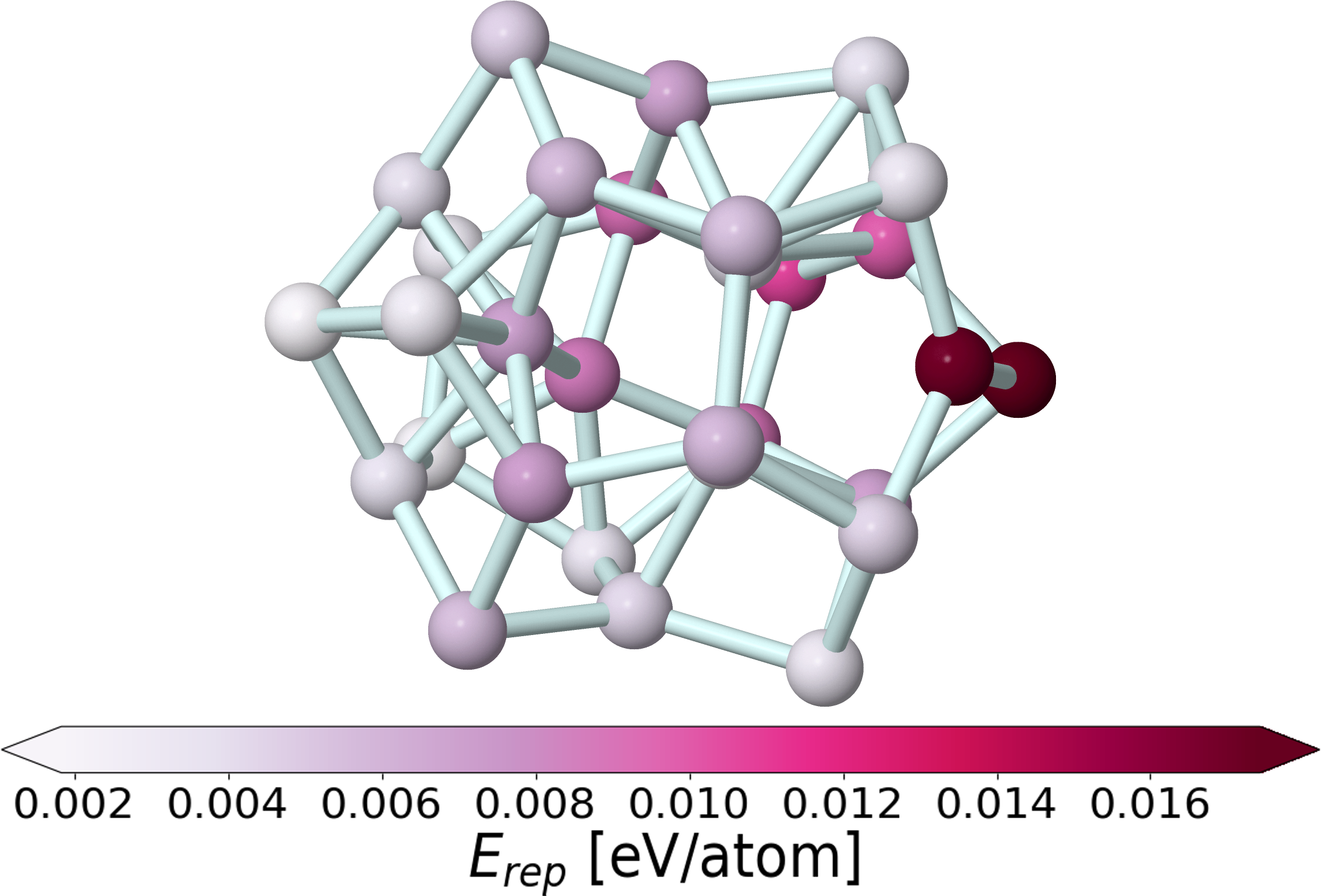}\\ \vspace{2ex}
    {(a) Compact $\text{Si}_{30}$ cluster optimized by Splines.} 
    \vspace{2ex}
  \end{minipage} \hfill
  \begin{minipage}[b]{0.49\linewidth}
    \centering
    \includegraphics[width=\textwidth]{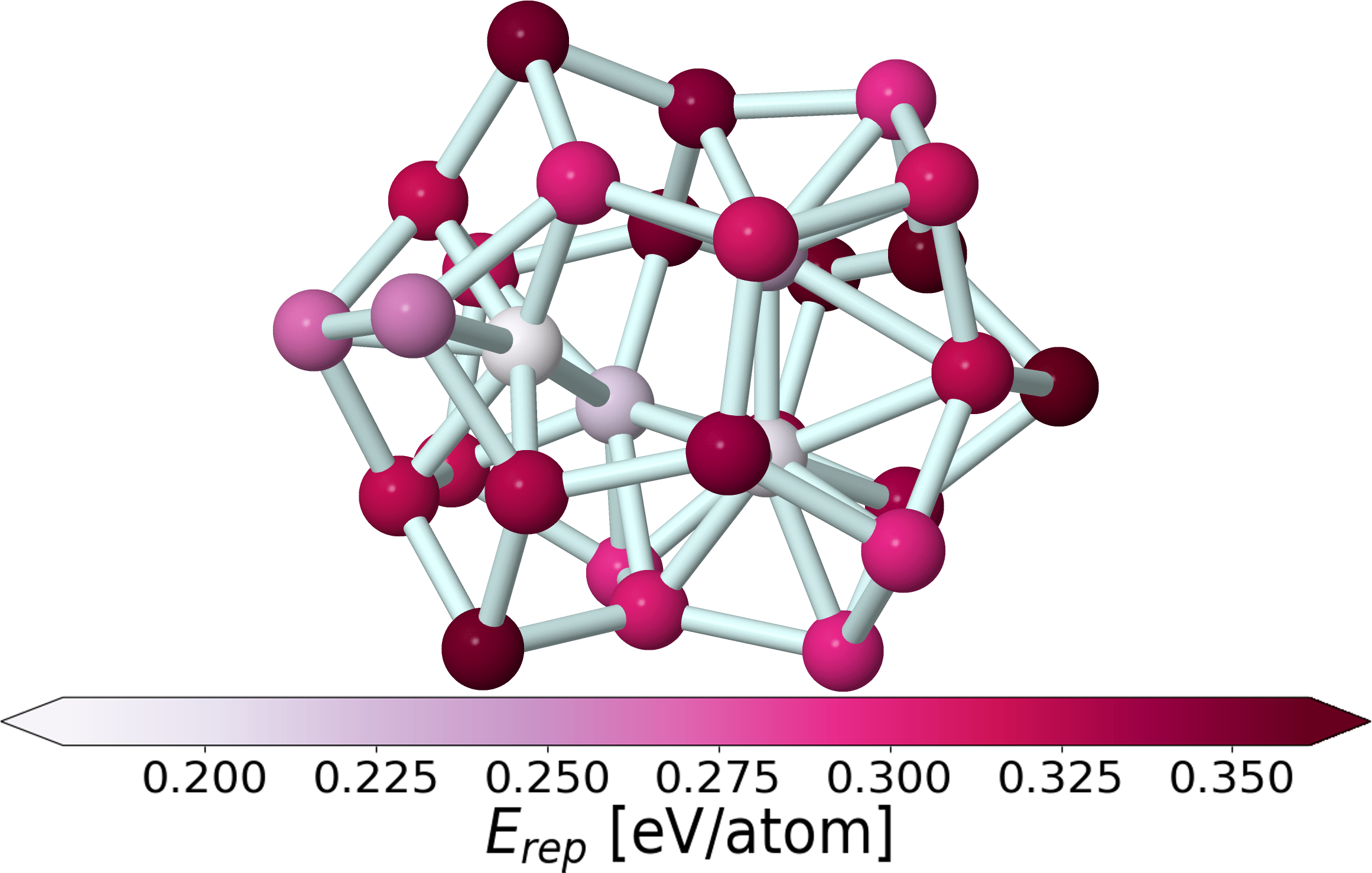}\\ \vspace{2ex}
    {(b) Compact $\text{Si}_{30}$ cluster optimized by NN-E$_\text{rep}$ .}
    \vspace{2ex}
  \end{minipage} \\
  \begin{minipage}[b]{0.49\linewidth}
    \centering
    \includegraphics[width=\textwidth]{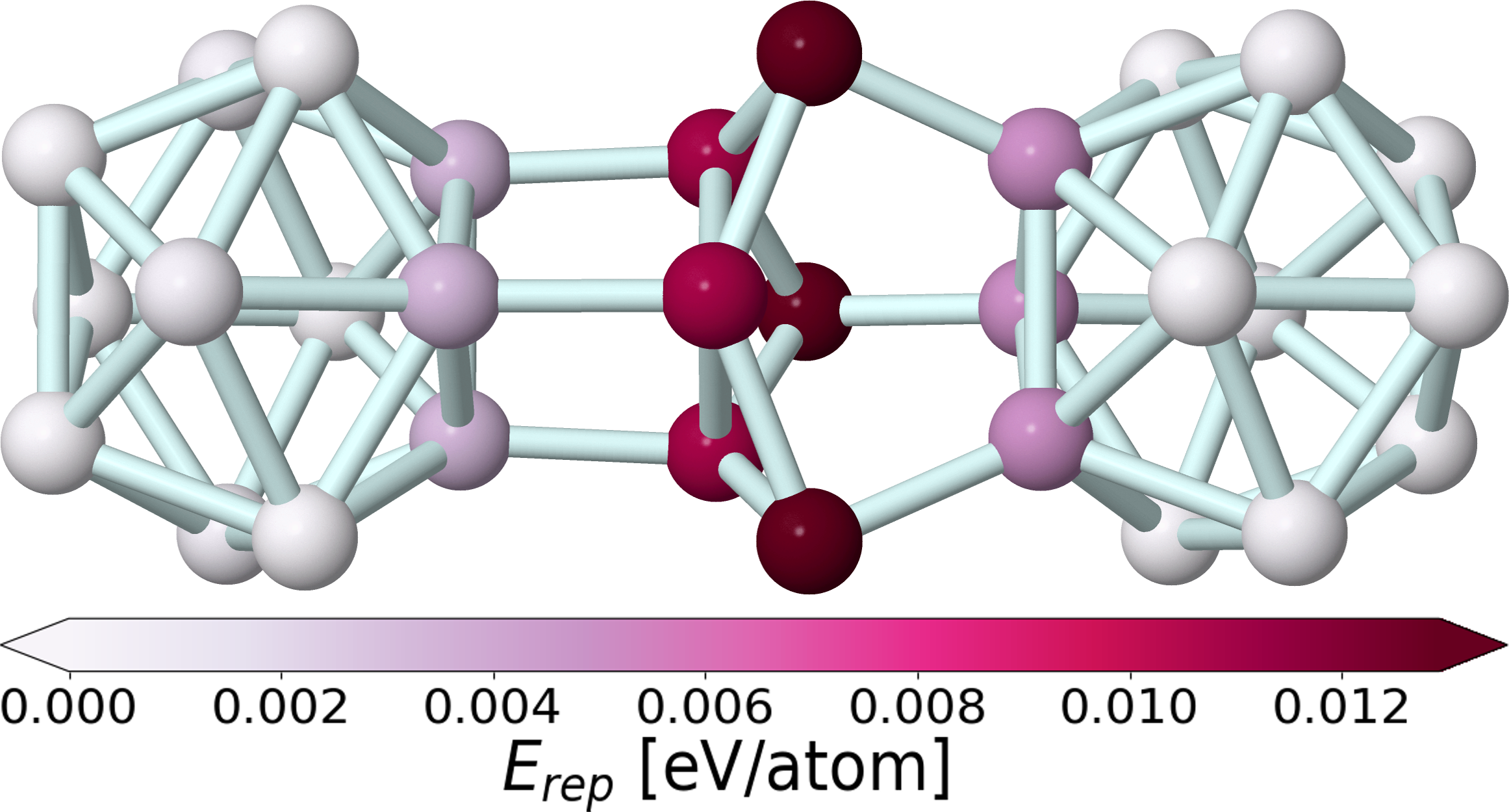}\\ \vspace{2ex}
    {(c) Elongated $\text{Si}_{30}$ cluster optimized by splines.} 
  \end{minipage} \hfill
  \begin{minipage}[b]{0.49\linewidth}
    \centering
    \includegraphics[width=\textwidth]{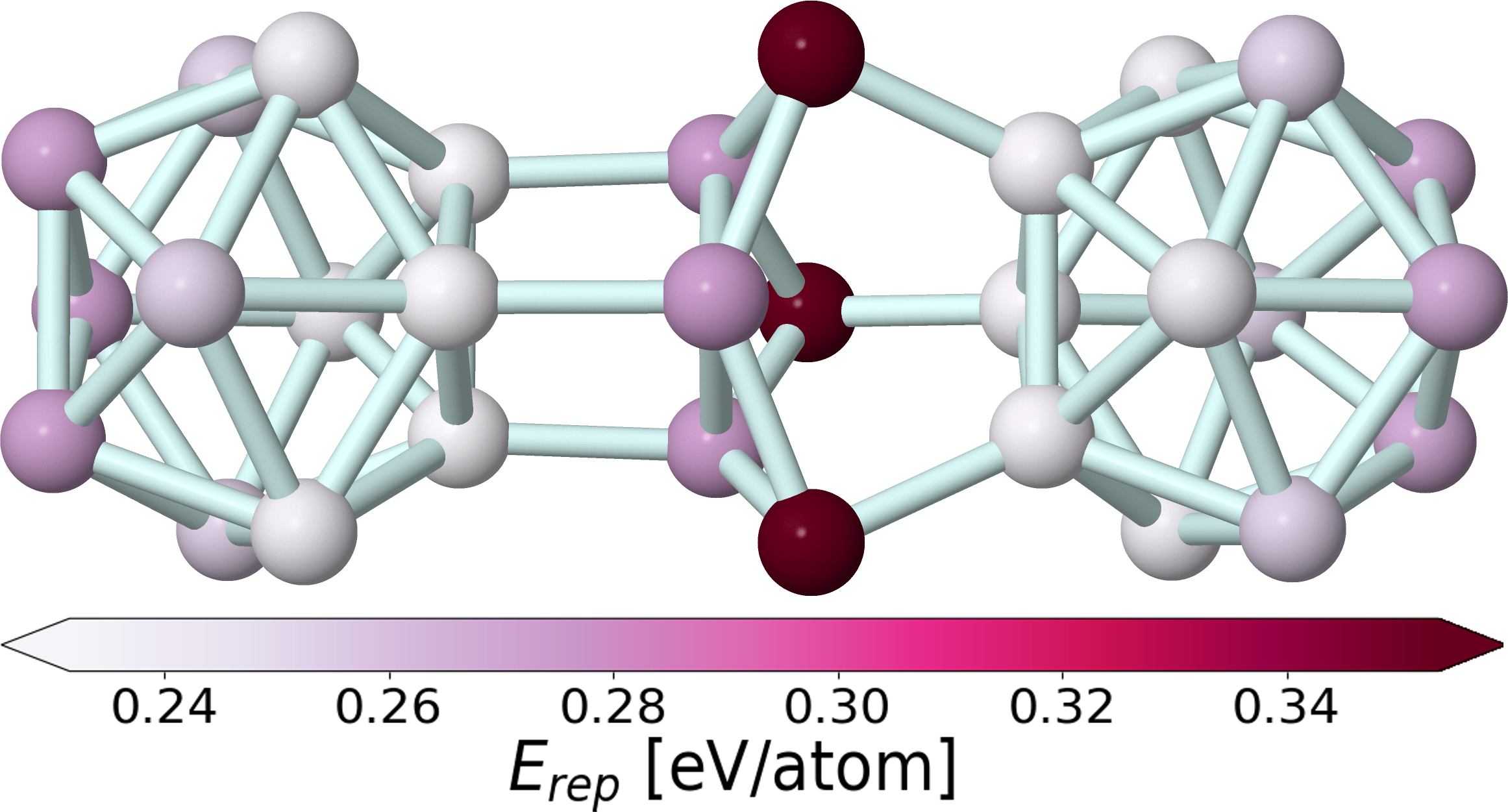} \\ \vspace{2ex}
    {(d) Elongated $\text{Si}_{30}$ cluster optimized by NN-E$_\text{rep}$ .} 
  \end{minipage} 
  \caption{\label{fig:Si30_Erep_atom} Repulsive energies per atom for two configurations of $\text{Si}_{30}$ nano-clusters predicted by Splines and NN-E$_\text{rep}$ .}
\end{figure}

While most cluster sizes are well represented by both repulsive potentials, Splines are overall the most robust of the two, and are hence more reliable despite a tendency to overestimate $r_g$. NN-E$_\text{rep}$ sporadically exhibits large errors. In particular, we notice a major disagreement for $30$ atoms.
We present in figure \ref{fig:Si30_Erep_atom} repulsive energies per atom predicted by both repulsive potentials for two $\text{Si}_{30}$ configurations. The first configuration, more compact, is predicted as the most stable by Splines, in agreement with DFT/VASP. The second one, with an elongated shape, is predicted as the most stable by NN-E$_\text{rep}$.

There is a noticeable difference in the handling of tetrahedra on the cluster surface. A detailed look at each cluster shows that both methods agree on the geometry with similar angles and distances. These tetrahedra have Si-Si-Si angles around $\sim 85^\circ$, while Si-Si distances vary from tetrahedron to tetrahedron. For NN-$E_\text{rep}$, 3-coordinated atoms on top of those tetrahedra (many on the surface in figure \ref{fig:Si30_Erep_atom}.b and in the center of figure \ref{fig:Si30_Erep_atom}.d) systematically exhibit a large $E_\text{rep}$. With Splines, atoms with high $E_\text{rep}$ (located on the far right of figure \ref{fig:Si30_Erep_atom}.a and in the middle of figure \ref{fig:Si30_Erep_atom}.c) belong to tetrahedra with Si-Si distances $\leq 2.35$ \AA, while similar geometries with larger distances have an almost vanishing $E_\text{rep}$. Unsurprisingly, Splines discriminate these configurations based on the interatomic distance only. In contrast, NN-$E_\text{rep}$ treats all tetrahedra with the same Si-Si-Si angles equally, regardless of distance, and clearly overestimates the repulsive energy. This points towards an underrepresentation of this configuration in the training process and highlights the difficulties one faces in generating balanced ML potentials.   

\section{\label{sec:computational_cost}Computational cost}
\begin{figure}[!htb]
  \centering
  \includegraphics[width=0.95\linewidth]{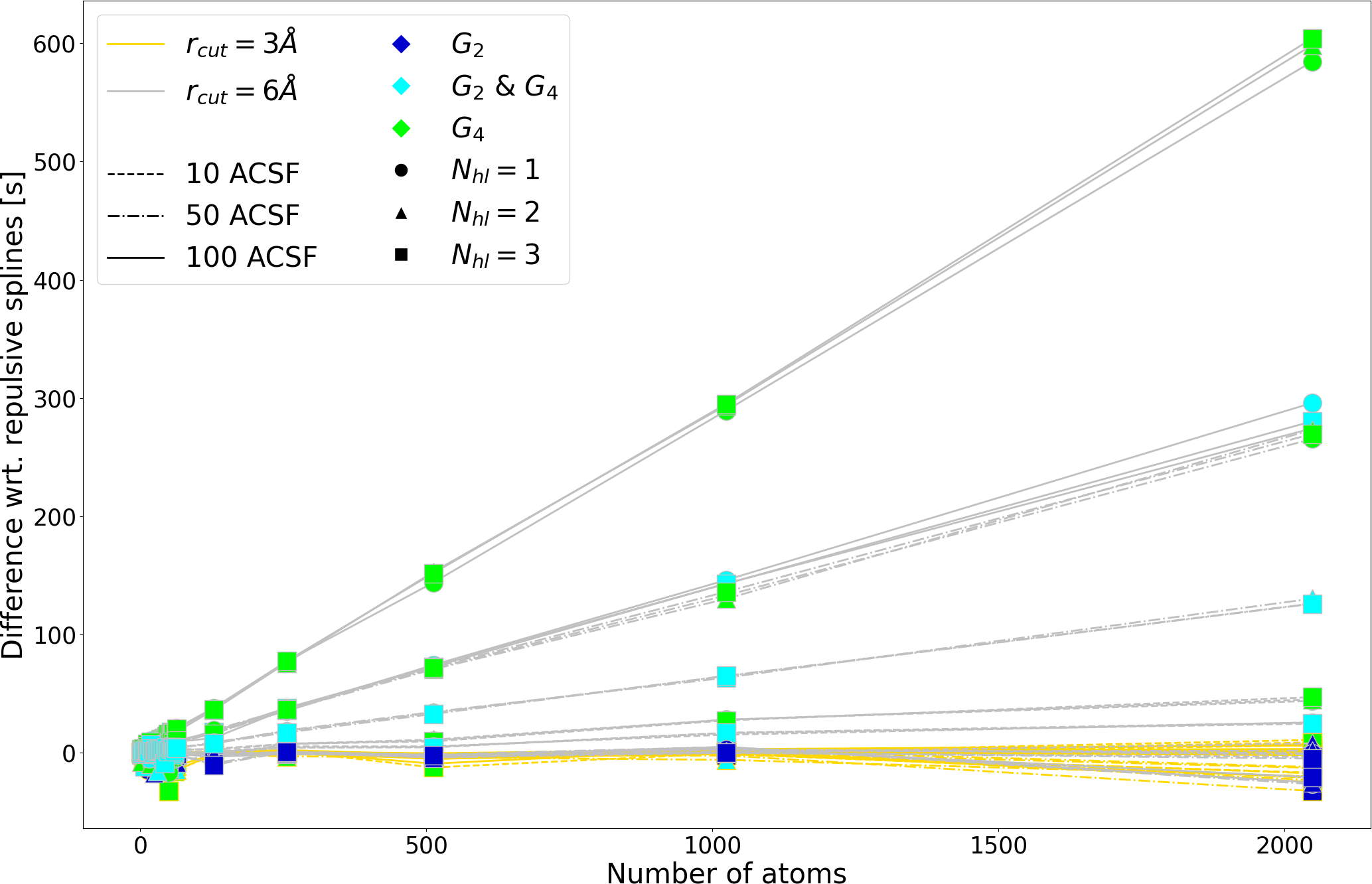}
  \caption{\label{fig:scaling_wrt_splines}Difference in CPU time of evaluating repulsive energies via various neural-networks and Splines for increasing sizes of ideal $\beta$-Sn systems.}
\end{figure}

In this last section we compare the computational scaling of Splines and neural-network repulsive potentials. The numerical cost of the latter is linked to several factors. First, the shape of the neural-network is relevant. We investigate network sizes of one to three hidden layers, and with input layers and hidden layers of either $\left\lbrace 10, 5 \right\rbrace$, $\left\lbrace 50, 25 \right\rbrace$ or $\left\lbrace 100, 50 \right\rbrace$ neurons respectively. Second, the type of symmetry functions can be of importance. We try networks with $G_2$ functions only, $G_4$ functions only, and a $50$\%-$50$\% mixture of both. Finally, the cutoff radius of the symmetry functions has an impact. We investigate $r_\text{cut}=3$ \AA\ (the same as used in this study) and $r_\text{cut} = 6$ \AA. For $\beta$-$Sn$ supercells containing various number of atoms we performed single point computations for each of those networks, as well as with the Splines and recorded the difference in computation time ($\Delta t = t_\text{CPU}^\text{NN} - t_\text{CPU}^\text{Spline}$). All computations were performed on a single CPU core (i7-6700K, 4.00GHz) and are reported in figure \ref{fig:scaling_wrt_splines}.

Unsurprisingly, the most important factor is the cutoff radius. For $r_\text{cut} = 6$ \AA\ computation times are significantly higher. The smaller cutoff ($r_\text{cut} = 3$ \AA) is close to the one used by Splines ($r_\text{cut} = 2.54$ \AA) and exhibits the same amount of neighbours ($6$ per atom) for $\beta$-$Sn$. In some cases, a large number of $G_4$ functions leads to an increase of the computation time. For a smaller cutoff (i.e. when Splines and neural-networks have the same number of neighbours) no particular performance relevant factor can be singled out. 
For the same number of neighbours per atom, the methodology used here has a similar computational effort as Splines. The time spent computing $E_\text{rep}$ with a neural-network of cutoff $3$ \AA\ never exceeds $1$ \% of the total computation time for the systems studied here. The numerical scaling of NNs is slightly worse than splines, but as it scales much better than the matrix diagonalization needed for the electronic part of the energy, the relative cost of $E_\text{rep}$ decreases as systems get larger.

\section{\label{sec:conclusion}Summary}
The main interest of adopting a many-body representation for the DFTB repulsive energy is the ability to optimize its parameters for a large set of reference data. This is achieved in two ways. The first one is based on the possibility to take into account any given system, as opposed to pair repulsive potentials which have to be trained on very specific (isolated or symmetric) systems. The second is the flexibility to be trained on a large variety of systems, whereas even the most refined splines representation cannot be generalized to several systems with very similar bond lengths\cite{bib:akak_css_hdnnp}. The observations in the present article partially agree with this assertion. Neural networks lead to an improvement of varying degree for energy vs. volume curves, formation energies of interstitials, angular distribution functions of liquids and amorphous systems. The improvement is also noticeable, although less significant, on other properties like the VDOS for bulk systems or the radial distribution functions of liquids at temperatures not exceeding the ones used in the training. For higher temperatures, extrapolation occurs and is poorly handled. The representation of clusters is also altered by the neural-network, with a noisy potential energy surface due to the discontinuous form of the many-body potential. The performance gain (or loss) brought upon by the ML repulsive potential is restricted by the intrinsic accuracy of the electronic terms of the DFTB energy. 

This is even more noticeable when comparing NN-E$_\text{rep}$ to full-ML potentials (GAP/NNP) which are directly fitted to DFT. While the latter make the most out of their flexibility, this comes as a double-edged sword. Properties for structures close to the training data are given with higher accuracy (cf. liquids), far from the training data unphysical results might occur (cf. spurious minima in the E(V) data, discontinous PES). Fitting only the correctional term that is the repulsive energy of DFTB allows the other terms to act as safety nets, with the additional benefit of providing access to the electronic states. The initial hope that these terms would help to bridge between finite clusters and solids by taking long-range interactions naturally into account were however not completely fulfilled. Results for the clusters show that NN-E$_\text{rep}$ indeed outperforms full-ML potentials, but is not better than the simple Spline representation. 

Therefore, it is safe to conclude that while machine-learned repulsive potentials can offer a clear improvement over pair potentials, it comes with the necessity to be vigilant in the choice of reference data and the properties to be represented. Carefully constructed splines are, by contrast, reliable and assure at least a robust physical prediction for any given system. The quality of the other DFTB energy contributions is also of key importance, as even highly accurate and flexible short range potentials seem unable to fully compensate for their deficiencies.

\begin{acknowledgments}
We thank GENCI for computational resources under projects DARI A0050810637 and A0070810637.
\end{acknowledgments}

\bibliography{bibliography}

\end{document}